\documentclass[lettersize,journal]{IEEEtran}
\usepackage[caption=false,font=normalsize,labelfont=sf,textfont=sf]{subfig}
\usepackage{url}

\usepackage{amsmath,amsfonts}
\usepackage{array}
\usepackage{textcomp}
\usepackage{stfloats}
\usepackage{verbatim}
\usepackage{graphicx}
\usepackage{booktabs}                      
\usepackage{hyperref}
\usepackage{algorithmic}
\usepackage[table,xcdraw]{xcolor} 
\usepackage[T1]{fontenc}    
\usepackage{amsfonts}       
\usepackage{nicefrac}       
\usepackage{microtype}      
\usepackage{cleveref}
\usepackage[ruled,linesnumbered]{algorithm2e}
\usepackage{multirow}

\usepackage{amssymb}
\usepackage{amsfonts}
\usepackage{makecell}
\usepackage{multirow}
\usepackage{pifont}
\usepackage{bm}

\def\1{{\bf{1}}}
\def\0{{\bf{0}}}

\newtheorem{definition}{Definition}

\def\BibTeX{{\rm B\kern-.05em{\sc i\kern-.025em b}\kern-.08em
    T\kern-.1667em\lower.7ex\hbox{E}\kern-.125emX}}
\usepackage{balance}
\begin{document}
\title{Identifying Evidence Subgraphs for Financial Risk Detection via Graph Counterfactual and Factual Reasoning}
\author{Huaming Du, Lei Yuan, Qing Yang, Xingyan Chen, Yu Zhao, Han Ji, Fuzhen Zhuang, Carl Yang, Gang Kou
\thanks{H. Du is with Financial Intelligence and Financial Engineering Key Laboratory of Sichuan Province, Southwestern University of Finance and Economics, China.

L. Yuan and G. Kou are with the School of Business Administration, Southwestern University of Finance and Economics, China. 

Q. Yang  is with Southwestern University of Finance and Economics, China.

X. Chen and Y. Zhao are with Financial Intelligence and Financial Engineering Key Laboratory of Sichuan Province, Institute of Digital Economy and Interdisciplinary Science Innovation, Southwestern University of Finance and Economics, China. 

H. Ji is with Ant Group, Beijing, China.

F. Zhuang is with Institute of Artificial Intelligence, SKLSDE, School of Computer Science, Beihang University, Beijing, China.

C. Yang is with Department of Computer Science, Emory University, Atlanta 30322 USA.

Corresponding author: Carl Yang (j.carlyang@emory.edu) and Gang Kou (kougang@swufe.edu.cn)}}

\markboth{Journal of \LaTeX\ Class Files,~Vol.~18, No.~9, September~2020}%
{How to Use the IEEEtran \LaTeX \ Templates}

\maketitle

\begin{abstract}
Company financial risks pose a significant threat to personal wealth and national economic stability, stimulating increasing attention towards the development of efficient and timely methods for monitoring them. Current approaches tend to use graph neural networks (GNNs) to model the momentum spillover effect of risks. However, due to the black-box nature of GNNs, 
these methods leave much to be improved for precise and reliable explanations towards company risks.
In this paper, we propose \textbf{CF\textsuperscript{3}}, a novel \underline{\textbf{C}}ounter\underline{\textbf{f}}actual and \underline{\textbf{F}}actual learning method for company \underline{\textbf{F}}inancial risk detection,  which generates evidence subgraphs on company
knowledge graphs to reliably detect and explain company financial risks. 
Specifically, we first propose a meta-path attribution process based on Granger causality, selecting the meta-paths most relevant to the target node labels to construct an attribution subgraph. 
Subsequently, we propose an edge-type-aware graph generator to identify important edges, and we also devise a layer-based feature masker to recognize crucial node features. Finally, we utilize counterfactual-factual reasoning and a loss function based on attribution subgraphs to jointly guide the learning of the graph generator and feature masker.
Extensive experiments on three real-world datasets demonstrate the superior performance of our method compared to state-of-the-art approaches in the field of financial risk detection.
\end{abstract}


\section{Introduction}
\IEEEPARstart{I}{n} the realm of finance, risk monitoring and identification, such as bankruptcy prediction \cite{liu2023qtiah}, credit scoring models \cite{kou2021bankruptcy}, and stock market forecasting \cite{zhao2023stock}, are of paramount importance. Especially in the era of ubiquitous connectivity, the interconnections between companies have become increasingly intricate, and companies encounter both internal risks and contagion risks \cite{wei2024combining}. Therefore, researching financial risks of companies is of paramount significance for individuals, government policymakers, and financial institutions. 

\begin{figure}[t]
	\centering
   \includegraphics[width=0.48\textwidth]{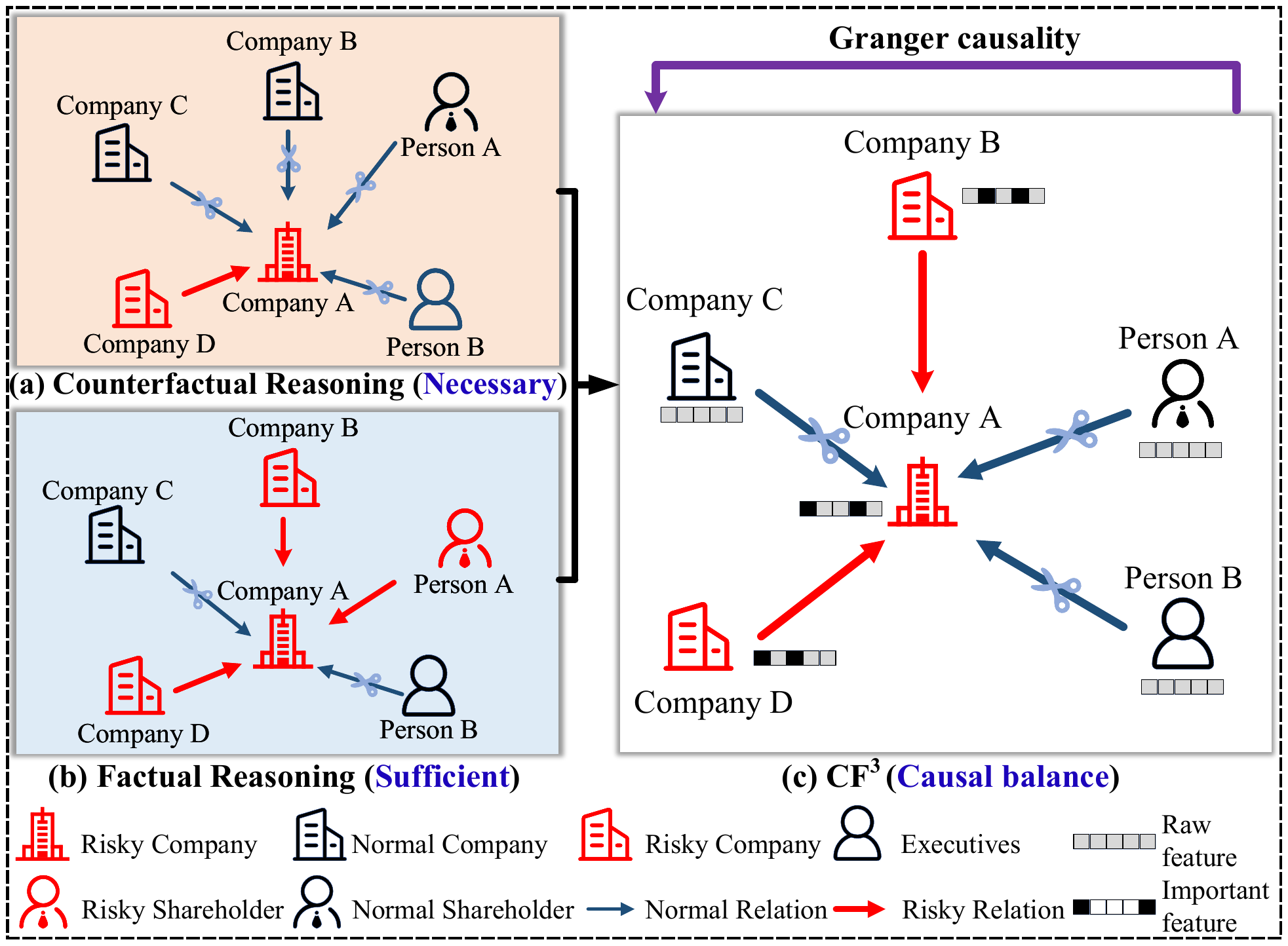}
	\caption{An example of extracting explanations for company financial risk detection. Based on the computational subgraph of target company node A, (a), (b), and (c) are the explanations extracted by counterfactual reasoning, factual reasoning, and CF\textsuperscript{3}, respectively. The subgraph in (c) also represents the ground-truth explanation.}
	\label{figure1}
\end{figure}  

With technological advancements in natural language processing (NLP) and AI, researchers have begun to evaluate company financial risk from a Big Data perspective. Besides traditional company financial index, more company risk intelligence is taken into consideration, such as non-financial
textual information and relational data. Specifically, to deal with textual risk information, NLP techniques, including sentiment analysis \cite{zhang2022survey} and event extraction \cite{huang2023event} are used to dig company risk signals from non-financial textual data. To model the risk momentum spillover\footnote{There is a risk momentum spillover effect among related companies, where the risk of one company may lead to the risk of other companies associated with it.} on company relational data \cite{xu2021rest,cheng2021modeling,bi2022company,liu2023qtiah,wang2023financial}, AI techniques, including deep learning and Graph Neural Networks (GNNs), are utilized to evaluate the company contagion risk. This series of studies have opened up new avenues for analyzing the generation and contagion mechanisms of company financial risk. Despite the successful research in company financial risk prediction, most existing  methods \cite{cho2023feature,nguyen2023bankruptcy,schnake2021higher,li2023pen,koa2024learning,yu2024fusing,wang2025risk} have failed to explain ``why does the model make such predictions? and which features in the input sample caused the sample label?'', i.e., lacking of interpretability. 

To address this challenge, we propose \textbf{CF\textsuperscript{3}}, a novel interpretation method for company financial risk detection by extracting evidence subgraphs\footnote{The evidence subgraph consists of crucial graph structures and node features that determine the model's output. As shown in Figure \ref{figure1}(c), nodes A, B, and D, along with their corresponding partial node features, form the evidence subgraph for risk entity A.}. Different from existing methods \cite{cho2023feature,li2023pen,koa2024learning}, we use counterfactual-factual reasoning theory to derive causal balance explanations. As shown in Figure \ref{figure1}, explanations obtained through factual reasoning are sufficient but not necessary, and may contain redundant information (e.g., Person A); explanations derived from counterfactual reasoning are necessary but not sufficient, and may only extract a small subset of the true explanation, potentially missing important risk information (e.g., Company B). Unlike existing methods \cite{nguyen2023bankruptcy,li2023pen,koa2024learning}, we have also considered the synergistic effect between graph structure and node features. By addressing these two aspects, we have obtained more accurate and reliable explanations, as shown in Figure \ref{figure1}(c). 
Specifically, adhering to the Granger causality principle \cite{granger1980testing}, we first select crucial meta-paths through a meta-path-based attribution process.  
Subsequently, to identify important edges under different company relationships, we employ an edge-type graph generator for graph reconstruction. For node features, we introduce a layer-based masking matrix designed to modify the node feature space. Finally, we utilize counterfactual-factual reasoning and a loss function based on attribution subgraph to extract significant subgraph structures and key node features for monitoring company financial risk.

CF\textsuperscript{3} differs from traditional subgraph-level GNN explainability methods \cite{lin2021generative,lueig2023,lugoat2023,azzolin2023global,rong2024efficient,lu2024eig,armgaan2024graphtrail,chen2024interpretable} in the following ways. \textit{First}, CF\textsuperscript{3} takes into account the synergistic effects of graph structure and node features on company financial risk. \textit{Second}, CF\textsuperscript{3} focuses on the causal relationships between substructures, features, and model outputs. \textit{Finally}, our method extracts causally-related evidence subgraphs by synergizing graph structure and node features from a global perspective, rather than focusing on single-instance explanations. This enables the identification of company risk sources and their propagation paths, facilitating timely risk warning and response, thereby reducing systemic risk.

The contributions of this paper are summarized as follows:
\begin{itemize}
    \item To the best of our knowledge, this is the first explanatory method based on counterfactual-factual reasoning for financial risk detection.
   \item We leverage Granger causality theory to propose an attribution process based on meta-paths, a graph generator based on multiple relationship types, and a layer-wise feature masker based on nonlinear mapping to obtain evidence subgraphs.
  \item We conduct extensive experiments on three real-world company bankruptcy datasets. And the results demonstrate superiority of the proposed model over the 13 SOTA baselines.
\end{itemize}
\section{Related Work}
Our work is related to financial risk prediction, interpretability of risk prediction, and counterfactual reasoning mechanism.
\subsection{Company Financial Risk Prediction}
Company financial risk prediction aims to forecast whether a company is exposed to a specific type of risk by utilizing various company-related information.
The main types of financial risks for businesses include credit risk \cite{tang2010market}, bankruptcy risk \cite{zheng2021heterogeneous,biddle2022accounting}, and guarantee risk \cite{cheng2020contagious}. At the data level, many studies utilize financial index \cite{Khashman2010NeuralNF,Cheng2022Regulating}, non-financial textual information \cite{Hu2020Loan,ye2020financial}, relational data \cite{bi2022company,wang2021ignorance}, and intelligence integration \cite{gofman2022trade,zhao2023stock} to predict company financial risk. At the model level, there are many studies on improving the prediction performance via statistical model \cite{xu2021internet}, machine learning model \cite{tobback2017bankruptcy}, deep learning model \cite{Yin2020Evaluating,Yang2020Financial} and hybrid model \cite{chuang2013application}.
Compared to studies focused on the performance of risk prediction models, only a few studies have focused on the interpretability problem of the machine learning-based models for tabular data \cite{cho2023feature,nguyen2023bankruptcy}.

\subsection{Interpretability of Risk Prediction}
Despite significant advances in deep company risk prediction models, their application remains highly limited due to their "black box" nature. To date, only a few research papers have addressed the issue of prediction interpretability \cite{hu2018listening,dang2021squawk}. Previous work includes Li \textit{et al.} \cite{li2023pen} proposed a Shared Representation Learning (SRL) module to explain the importance of various text documents within a text corpus. More recently, Koa \textit{et al.} \cite{koa2024learning} introduced a Summary-Explanation-Prediction (SEP) framework that fine-tunes a specialized large language model to autonomously learn how to generate interpretable stock predictions. However, these methods focus solely on text data and do not adequately model the risk spillover effects between companies. 

Some recent studies have attempted to use GNNs to capture relationships between companies. However, due to the complexity of financial scenarios, recent SOTA GNN explanation methods, which are based on either factual reasoning \cite{yuan2021explainability,huang2022graphlime,zhang2022protgnn,zhang2022gstarx,lueig2023,lugoat2023,chen2024interpretable} or counterfactual reasoning \cite{abrate2021counterfactual,bajaj2021robust,lucic2022cf,chen2024interpretable}, as well as post-hoc interpretation methods \cite{ying2019gnnexplainer,luo2020parameterized} and inherently interpretable models \cite{miao2022interpretable,miao2023interpretable,wu2022discovering}, can not be directly applied to company financial risk detection. The main reasons are \underline{as follows}: \textit{First}. Traditional GNN interpretability methods focus on explaining either graph structures or node features, but in finance, both internal company risk (node features) and contagion risk (graph structure) are essential. \textit{Second}. These methods model correlations without fully capturing causal relationships, limiting their ability to characterize complex financial risks. \textit{Finally}. They emphasize individual subgraph interpretability but lack the capability to model the universality of financial risks. In contrast, our method synergizes graph structure and node features, extracting causally-related evidence subgraphs from a global perspective. This helps capture sources of company risk and their propagation paths, reducing systemic risk.
\subsection{Counterfactual Reasoning Mechanism}
Counterfactual reasoning is a basic way of reasoning that helps people understand their behaviours and the world’s rules \cite{li2020survey}. A counterfactual explanation is an effective type of example-based local explanation that offers changes needed to produce a different outcome from the model for a specific instance. 
Recently, some graph counterfactual explanation methods have been proposed \cite{abrate2021counterfactual,tan2022learning,lucic2022cf,huang2023global}. 
Taking Figure \ref{figure1}(a) as an example, counterfactual reasoning generates a subgraph with only two edges. If these edges are removed, it indeed leads to a different prediction (i.e., normal company). However, such an explanation does not encompass the complete risk information about the target company. 



\section{PRELIMINARIES}
We are given a \textit{company
knowledge graph} $G=(\mathcal{V},\mathcal{E})$, where $\mathcal{V}$ represents the company and stakeholders, $\mathcal{E}$ represents multiple types of edges between nodes, and $X \in \mathbb{R}^{n \times d}$ represents a set of all node features. $n$ is the number of nodes and $d$ represents the dimension of node features. 

\begin{definition}
    \textbf{(Meta-path)} A meta-path $P$ is defined as a path
in the form of $O_{1}\xrightarrow{r_{1}}O_{2}\xrightarrow{r_{2}} \cdots \xrightarrow{r_{l-1}} O_{l}$ which describes a composite
relation $R = r_{1} \circ  r_{2} \cdots r_{l-1}$ between node types $O_{1}$ and $O_{l}$
, where $\circ$ denotes the composition operator on relations.
\end{definition}
For example, $E_{1}\xrightarrow{investment}E_{2}\xrightarrow{supply} E_{3}$ is a meta-path sample of
meta-path $E\xrightarrow{investment}E\xrightarrow{supply} E$.


\begin{definition}
    \textbf{(Granger causality)} Granger causality describes the relationships between two (or more) variables, where one variable causes a change in another. Specifically, if we can improve the prediction of the variable $\hat{y}$ using all available information $U$ compared to using information excluding the variable $x_{i}$, we say that $x_{i}$ Granger causes $\hat{y}$ \cite{granger1980testing}, denoted as $x_{i}\to \hat{y}$.
\end{definition}
Financial risk detection task can be formulated as follows:
\begin{definition}
    \textbf{(Company financial risk detection)
    } In company konwledge graph $G$, the target company $v_{i}\in \mathcal{V} $ has a label $y_{i}$, and the L-hop computation subgraph of $v_{i}$ is defined as $\mathcal{G}_{s}=\left ( \mathcal{V}_{s},\boldsymbol{A}_{s},\boldsymbol{X}_{s}\right )$.
The detection task aims to generate the explanation for risk prediction as evidence subgraph $G_s =(\boldsymbol{\hat{A}},\boldsymbol{\hat{X}})$
, which consists of \textbf{a subset of the edges} and \textbf{a subset of the feature space} of the computational graph $\mathcal{G}_{s}$.
\end{definition}
 
 Please note that company risk detection aims to capture evidence subgraphs that causally determine the model's predictions, while risk prediction aims to enhance the model's classification performance.

\section{METHODOLOGY}
\begin{figure*}[t]
	\centering
	\includegraphics[width=1\textwidth]{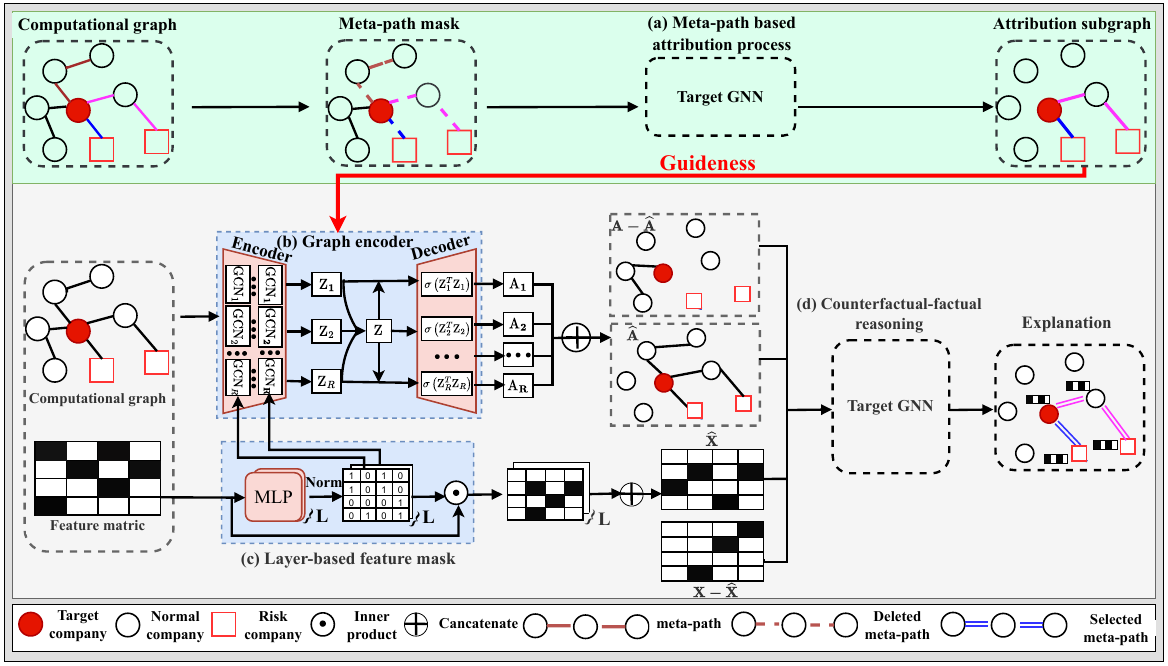}
	\caption{The overview of our proposed method: CF\textsuperscript{3}. \textit{(a) Meta-path based attribution process}: The aim is to generate the attribution subgraph for the target node by sequentially applying masks to each meta-path. \textit{(b) Graph generator}: Designed with an encoder-decoder architecture, this module outputs reconstructed subgraph structures for the target node under different relationship networks. \textit{(c) Layer-based feature mask}: Intended to produce feature mask matrices for different GCN layers and a global feature mask matrix. \textit{(d) Counterfactual and factual reasoning}: The objective is to infer important subgraph structures and crucial node features through reasoning based on factual and counterfactual scenarios. The target GNN (such as HAT \cite{zheng2021heterogeneous}, DANSMP \cite{zhao2022stock}, etc., used in the experimental section of this paper) is
pre-trained, and the parameters would not be changed during the training of CF\textsuperscript{3}. \textit{Please note that once training is complete, CF\textsuperscript{3} can utilize the (b) and (c) to construct explanations for the target GNN with minimal time consumption}.}
	\label{framework}
\end{figure*}

The overview of the proposed method is illustrated in Figure \ref{framework}. This section will give a detailed illustration of the proposed method. 

\subsection{Meta-path based  Attribution Process}\label{attributuion process}
In general, internal and external risks of a company often propagate and diffuse along the network of company relationships, potentially affecting the entire industry or market adversely. In heterogeneous company relationship networks, meta-paths possess rich and practical semantics, allowing for the modeling of microstructures and intrinsic operational mechanisms within the market. Inspired by existing research \cite{wang2019heterogeneous,zhang2023page}, we opt to model company financial risks using meta-path paradigms, as opposed to considering only pairwise relationships \cite{lin2021generative}. Specifically, we integrate Granger causality testing to quantify the causal contributions of meta-paths to the output of the predictive model. 

Given a target GNN and an instance $\mathcal{G}_{s}$, we use
$\delta _{\mathcal{G}_{s}}$ to denote the model error of the target GNN when considering the computation graph, while $\delta _{\mathcal{G}_{s} \backslash p}$ represents the model error excluding the information from the meta-path $p$. With these two definitions and following the notion of Granger causality, we can quantify the causal contribution of a meta-path $p$ to the output of the target GNN. More specifically, the causal contribution of the meta-path $p$ is
defined as the decrease in model error \footnote{In the attribution process, we intervene on specific meta-paths $p$ and consider the model error $Y$ under both the presence $p=1$ and absence $p=0$ of edges. Based on the potential outcomes framework theory, we can calculate the causal effect $Y\left (p=1 \right )-Y\left (p=0 \right )$ of a particular meta-path on the model prediction. Therefore, causal contribution is built upon error contribution, with causal contribution focusing more on causality, while error contribution emphasizes statistical correlation, which is consistent with existing research \cite{lin2021generative}.}, formulated as Eq. \ref{dalta}:
\begin{equation}
\label{dalta}
    \Delta _{\delta,p }=\delta _{\mathcal{G}_{s} \backslash p}-\delta _{\mathcal{G}_{s}}.
\end{equation}
To calculate $\delta _{\mathcal{G}_{s}}$ and $\delta _{\mathcal{G}_{s} \backslash p}$, we first compute the outputs
corresponding to the computation graph $\mathcal{G}_{s}$
and the one
excluding meta-path $p$ , $\mathcal{G}_{s} \backslash p$, based on the target GNN $f\left ( \cdotp \right )$. Then, the associated outputs can be formulated as follow:
\begin{equation}
\label{dalta1}
    \hat{y}_{\mathcal{G}_{s}}=f\left ( \mathcal{G}_{s} \right ), \quad \hat{y}_{\mathcal{G}_{s} \backslash p}=f\left (\mathcal{G}_{s} \backslash p \right ).
\end{equation}
Subsequently, we compare the disparity between the output of the target GNN and the ground-truth labels. Here, we use the loss function $\mathcal{L}$ of the target GNN
as the metric to measure the model error. The formulations for $\delta _{\mathcal{G}_{s}}$ and $\delta _{\mathcal{G}_{s} \backslash p}$ are outlined as follows:
\begin{equation}
\label{dalta2}
   \delta _{\mathcal{G}_{s}}=\mathcal{L}\left ( y,\hat{y}_{\mathcal{G}_{s}}\right ),  \quad   \delta _{\mathcal{G}_{s} \backslash p}=\mathcal{L}\left ( y,\hat{y}_{\mathcal{G}_{s} \backslash p}\right ).
\end{equation}
Now, the causal contribution of a meta-path $p$ can be measured by the loss difference associated with the computation graph and the one deleting meta-path $p$. Essentially, $\Delta _{\delta,p }$ can be regarded as capturing the Individual Causal Effect (ICE) of the input $\mathcal{G}_{s}$ with the value $p$ on the output $\hat{y}$ \cite{goldstein2015peeking}, serving as a metric to quantify the importance of the meta-path in risk prediction. Therefore, based on $\Delta _{\delta,p }$, we can directly obtain the most relevant subgraph for predicting $\hat{y}$, which we refer to as the ground-truth attribution process. The subgraph $G_{att}$ obtained from this attribution process can be utilized as one of the constraints in training our graph generator, aiming to encourage its outcomes to be effective explanations. 

\subsection{Graph Generator}\label{graph generator}
Current studies typically perform random perturbations on the topological structure to obtain local explanations for individual instances \cite{ying2019gnnexplainer,tan2022learning}. We believe that random perturbations cannot yield consistent explanatory results, failing to ensure model convergence, and, more importantly, failing to provide global explanations for similar samples \cite{wang2021towards,yang2023counterfactual}. 

We focus on utilizing a graph generator composed of a graph convolutional network encoder and an inner product decoder \cite{kipf2016variational} as the causal explanation model. More specifically, in our explainer, we initially decompose the subgraph of the target node into multiple relation graphs based on the types of edges. We apply graph convolutional layers to aggregate neighborhood information and learn node features for each relation graph. Subsequently, we use an inner product decoder to generate an adjacency matrix $A_{r}$ with respect to relation $r$, representing the important substructure that contains the most relevant portion of the computation graph for its prediction. Finally, we combine the substructures from different relations to obtain substructures for multiple relations. Particularly, each value in $A_{r}$ signifies the contribution of a specific edge to the prediction of $\mathcal{G}_{s}$, given that the edge exists in the target instance. Formally, the encoding process for a specific relation $r$ can be expressed as:
\begin{equation}
    \boldsymbol{Z}_{r}=\sigma\left ( \boldsymbol{A}_{s}\sigma \left ( \boldsymbol{A}_{s}\boldsymbol{X}_{s}\boldsymbol{X}_{m0}\boldsymbol{W}_{0}\right )\boldsymbol{X}_{m1}\boldsymbol{W}_{1}\right ) \ , 
\end{equation}
where $\boldsymbol{Z}_{r}$ is the learned node features, $\boldsymbol{A}_{s}$ is the adjacency matrix of the computation graph for target instance, $\boldsymbol{X}_{s}$ is the node features, $\boldsymbol{W}_{0}$ and $\boldsymbol{W}_{1}$ are the learnable parameters for the first and second layers of the GNN\footnote{To illustrate more clearly, we assume that the target model consists of two layers of GNN.}, respectively. $\boldsymbol{X}_{m0}$ and $\boldsymbol{X}_{m1}$ denote the feature mask matrices for the respective layers, which can be calculated using Eq. \ref{norm}.

The node embeddings learned under each type of company relationship signify specific local company affiliations, such as supply-demand relationships. In order to amalgamate the entire company relationship network, enabling the company embeddings to capture risk factors in both local and overarching environments, our encoding process involves initially obtaining global company embeddings. Subsequently, both global and local company embeddings are utilized as inputs to the decoder for the identification of crucial company relationships. The global embedding of the target company can be calculated as follows:
\begin{equation}
    \boldsymbol{Z}= \boldsymbol{Z}_{1}\oplus\boldsymbol{Z}_{2}\oplus \cdots \oplus\boldsymbol{Z}_{R} \ ,
\end{equation}
where $\boldsymbol{Z}$ is the global embedding, $R$ represents the set of company relationships, and $\oplus$ denotes addition operation. 

Through the decoder, the low-dimensional representation obtained from the encoder is reconstructed to highlight significant subgraph structures. In order to comprehensively model the importance of company relationships from different perspectives and obtain crucial subgraphs under specific relationships, we concatenate the low-dimensional representation ${Z}_{r}$ from the encoder under a particular relationship with the global embedding. This concatenated vector serves as the input to the decoder, enabling the extraction of important subgraph structures from both global and local perspectives of company relationships. The decoding process for a specific relation $r$ can be expressed as:
\begin{equation}
\label{generator1}
    \boldsymbol{Z}_{r}^{'}=\left [ \boldsymbol{Z}_{r}, \boldsymbol{Z}\right ] \ ,\quad     \boldsymbol{A}_{r}=\sigma \left (\boldsymbol{Z}_{r}^{'}{\boldsymbol{Z}_{r}^{'}}^\mathrm{T}\right ) \ ,
\end{equation}
where $\boldsymbol{Z}_{r}^{'}$ represents the low-dimensional representation obtained by incorporating the global embedding. $\boldsymbol{A}_{r}$ denotes the important subgraph structure obtained after incorporating the global embedding. We integrate the reconstructed subgraphs under different relationships to obtain the final reconstructed subgraph $\boldsymbol{\hat{A}}$.
\begin{equation}
\label{generator2}
    \boldsymbol{\hat{A}}= \boldsymbol{A}_{1}\oplus\boldsymbol{A}_{2}\oplus\cdots \oplus\boldsymbol{A}_{R} \ ,
\end{equation}
\subsection{Layer-based Feature Masker}\label{feature mask}
Node features are essential attributes of the nodes themselves, playing a crucial role in node classification tasks \cite{xiao2023counterfactual}, particularly in the context of company financial risk prediction \cite{cheng2021modeling}. The design of feature masking aims to perturb node features, simulating changes in the intrinsic attributes of the company, and exploring its internal risks. Existing research often interprets feature importance from an individual perspective, which can result in inefficient explanations and significant potential for inconsistencies in interpretations \cite{ma2022clear}. Furthermore, within the same propagation path, risks brought about by neighbors of different orders vary. For instance, credit information from first-order neighbors may significantly impact the target node, while litigation information from second-order neighbors may also play a crucial role. Therefore, we propose a layer based feature mask and model the company's financial risk from a global perspective. This approach not only improves interpretability efficiency but also captures genuine risk factors for the company.

Firstly, we use MLP to map the feature matrix corresponding to the computed subgraph of the target node, aiming to capture common risk factors among companies through the mapped matrix. The expression is as follows:
\begin{equation}
\label{norm}
\boldsymbol{X}_{ml}=Norm\left ( \boldsymbol{M}_l  \boldsymbol{X}_{s}\right )\:\text{and}\:l\in \left [ 0,L\right ] \ ,
\end{equation}
    where $\boldsymbol{X}_{ml}$ is the feature mask matrix for the $l$-th layer of GCN, $L$ is the number of GCN layers, $Norm$ denotes the normalization operation, $\boldsymbol{M}_l$ is the corresponding  learnable parameters of MLP and $\boldsymbol{X}_{s}$ denotes the feature matrix corresponding to the computed subgraph of the target node. 

Secondly, the obtained layer-based mask matrix can be utilized as an input to the graph generator for graph reconstruction, as illustrated in the previous section. Moreover, to capture the global importance of node features, we integrate the obtained layer-based feature mask matrices to derive a global feature mask matrix. This global feature mask matrix is subsequently employed in subsequent counterfactual and factual reasoning.
\begin{equation}
\label{global-norm}
    \hat{\boldsymbol{X}}= \boldsymbol{X}_{s}\odot\boldsymbol{X}_{m1}\oplus \boldsymbol{X}_{s}\odot\boldsymbol{X}_{m2}\oplus \cdots \oplus\boldsymbol{X}_{s}\odot\boldsymbol{X}_{ml} \ ,
\end{equation}
where $\hat{\boldsymbol{X}}$ is the global features after masking, corresponding to the important node features of evidence subgraphs, and $\odot$ denotes the Hadamard product.


After performing the graph generator and layer-based feature masking, we can obtain the important graph structure $\boldsymbol{\hat{A}}$ and node features $\boldsymbol{\hat{X}}$. 
Through the aforementioned structure and feature coupling operations, the model can better understand the sensitivity and importance of subgraph structures and node features.

\subsection{End-to-end Model Training}
\subsubsection{Subgraph reconstruction error.}
We envision that the reconstructed matrix is a weighted adjacency matrix, reflecting the contributions of edges to its prediction. Now, we can apply the ``guidance" based on the concept of Granger causality to supervise the learning process. Specifically, we use the root mean square error between the reconstructed weighted matrices and the true causal contributions to guide model training.
\begin{equation}
    \mathcal{L}_{recon}=MSE\left ( \boldsymbol{\hat{A}}-\boldsymbol{A}_{att}\right ) \ ,
\end{equation}
where $\boldsymbol{A}_{att}$ is the adjacency matrix corresponding to $\boldsymbol{G}_{att}$. 
\subsubsection{Counterfactual and Factual reasoning.} Factual reasoning poses the question: ``If company A goes bankrupt, will company B go bankrupt as a result?" In contrast, counterfactual reasoning asks: ``If A does not go bankrupt, will B still go bankrupt?" In the context of company financial risk detection, factual reasoning generates risk sources that are consistent with the facts, resulting in the same prediction by the GNNs. Counterfactual reasoning generates risk sources that are inconsistent with the facts, leading to different predictions by the GNNs. In simple terms, factual reasoning seeks a set of edges/features that are sufficient to produce the same risk prediction as using the entire subgraph, while counterfactual reasoning seeks a necessary set of edges/features so that their removal would result in different risk predictions. Factual reasoning is as follows:
\begin{subequations}
\begin{align}
    \hat{y_{i}}=\Phi \left ( \boldsymbol{\hat{A}},\boldsymbol{\hat{X}}\right ) \label{Za} \ ,\\
    {y_{i}}'=\Phi \left ( \boldsymbol{\hat{A}},\boldsymbol{X}_{s}\right ) \label{Zb} \ ,\\
     {y_{i}}''=\Phi \left ( \boldsymbol{A}_{s},\boldsymbol{\hat{X}}\right )  \label{Zc} \ .
\end{align}
\end{subequations}
The original computational subgraph and node features can be divided into two parts for node label prediction: important and non-important, as suggested by \cite{lucic2022cf,zhang2023mixupexplainer}. In this way, we expect to find a subset to generate $\boldsymbol{\hat{A}}$ and $\boldsymbol{\hat{X}}$ which share the same prediction as using the whole computation graph while generating different prediction with the $\boldsymbol{A}_{s} \backslash \boldsymbol{\hat{A}}$ and $\boldsymbol{X}_{s} \backslash \boldsymbol{\hat{X}}$. Counterfactual reasoning is as follows:
\begin{equation}
    {y_{i}}'''=\Phi \left ( \boldsymbol{A}_{s} \backslash \boldsymbol{\hat{A}},\boldsymbol{X}_{s} \backslash \boldsymbol{\hat{X}}\right ) \ ,
\end{equation}
where $ \boldsymbol{A}_{s} \backslash \boldsymbol{\hat{A}}$ represents the complement of $\boldsymbol{A}_{s}$  with respect to $\boldsymbol{\hat{A}}$, and similarly, $\boldsymbol{X}_{s} \backslash \boldsymbol{\hat{X}}$ does the same. 

We simultaneously leverage subgraph reconstruction error, counterfactual and factual reasoning to guide the learning of the graph generator and feature masker. Therefore, the overall training loss of our proposed CF\textsuperscript{3} is:
\begin{equation}
\label{final}
  \begin{aligned}
    \mathcal{L}=&\alpha  \left ( CE\left ( y_{i},\hat{y_{i}}\right )+ CE\left ( y_{i},{y_{i}}'\right ) + CE\left ( y_{i},{y_{i}}''\right )\right )+ \\ &\beta  \mathcal{L}_{recon}+\gamma  CE\left ( 1-y_{i},{y_{i}}'''\right )+L_{reg} \ ,
   \end{aligned}
\end{equation}
where $CE\left (\ast \right )$ is the cross-entropy loss, $\alpha$, $\beta$, and $\gamma$ are the learnable parameters, and $L_{reg}$ is the $L_{2}$-regularizer of the model parameters. We establish the overall learning procedure as summarized in Algorithm \ref{algorithm1}.
\begin{algorithm}[tb]
\caption{Training the CF\textsuperscript{3} model}
\label{algorithm1}
\textbf{Input}: Company knowledge graph $G$, computation graph adjacency matrix $A_{s}$, computation graph feature matrix $X_{s}$, layers of GCN $L$. \\
\textbf{Output}: The important graph structure $\widehat{A}$, important node feature $\widehat{X}$.
\begin{algorithmic}[1] 
\STATE Initialize the trainable  parameter matrices and training epoch $I$.
\WHILE{$j\leqslant I$}
\STATE  \textbf{/* Layer-based Feature Masker */}
\STATE $\boldsymbol{X}_{ml}\leftarrow $ the feature mask operation for the $l$-th layer via Eq.(\ref{norm}).
\STATE $\hat{\boldsymbol{X}}\leftarrow $ the global feature mask operation via Eq.(\ref{global-norm}).
\STATE \textbf{/* Meta-path based Attribution Process */}
\STATE $G_{att}\leftarrow $ The meta-path based attribution process via Eq.(\ref{dalta},\ref{dalta1},\ref{dalta2})
\STATE \textbf{/* Graph Generator */}
\STATE  $\boldsymbol{A}_{r}\leftarrow $ the subgraph reconstruction under specific relationships via Eq.(\ref{generator1}).
\STATE $\boldsymbol{\hat{A}}\leftarrow $ the final subgraph reconstruction via Eq.(\ref{generator2}).
\STATE \textbf{/*End-to-end Model Training */}
\STATE Minimized the overall loss in Eq.(\ref{final}).
\ENDWHILE
\STATE \textbf{return} $\widehat{A}$, $\widehat{X}$.
\end{algorithmic}
\end{algorithm}

\noindent\textbf{Computational complexity analysis}.
Our approach mitigates the estimation cost by training both a graph generator and a feature masker, facilitating the generation of explanations for any given instances. Specifically, once the parameterized graph autoencoder and the feature masker within CF\textsuperscript{3} are trained, these components can be employed in an inductive setting to explain new instances. Importantly, the model parameter complexity of CF\textsuperscript{3} is independent of the input graph size. With the inductive property, the inference time complexity of CF\textsuperscript{3} is $\mathcal{O}\left ( R\left | E\right |\right )$, where $\left | E\right |$ represents the number of edges in the instance to be explained. 
Consequently, we can leverage trained graph generative models and mask models to quickly handle the explanation task.

\section{Experiments}
In this section, we conduct extensive experiments to answer the
following questions: \textbf{RQ1}: How does CF\textsuperscript{3} perform in company financial risk detection? \textbf{RQ2}: What is the impact of the different modules on the model performance? \textbf{RQ3}: What is the influence of hyper-parameter settings in the proposed model? \textbf{RQ4}: How is the computational performance of CF\textsuperscript{3}? \textbf{RQ5}: How does CF\textsuperscript{3} perform in real-world cases?

\subsection{Datasets}
To examine the performance of CF\textsuperscript{3} for company financial risk detection, HAT \cite{zheng2021heterogeneous}, DANSMP \cite{zhao2022stock} and ComRisk \cite{wei2024combining} are chosen as target-GNNs, and we choose three public and real-world graph datasets as shown in Table \ref{Dataset Analysis}: SMEsD \cite{wei2024combining}, CSI300E \cite{zhao2022stock}, and SME\footnote{Please note that other relevant datasets are not publicly available and cannot be used as experimental datasets for this article.} \cite{zheng2021heterogeneous}. The datasets statistics are summarized in \Cref{tab: Statistics,tab: Description}. 

\textbf{SMEsD:} For the company heterogeneous graph, there are four types of relationships between companies and persons and six types of relationships between companies and companies. 
For the hypergraph, there are three types of edges: industry, area, and stakeholder. We split SMEsD into a training set, validation set, and testing set across the bankruptcy time.
Table \ref{tab: Description} describes in detail the characteristics of each node, including basic business information, litigation events, and company credit information from 31 dimensions. Specifically, the business information includes registered capital, paid-up capital, and establishment time. Each litigation event includes the related plaintiff, defendant, subject, court level, result, and timestamp. Credit information includes information on trustworthiness incentives, administrative licenses, and business exceptions, where trustworthiness incentives include the year of evaluation, administrative licenses include the type and the time of license, and business exceptions include the date of establishment.
\begin{table}[t]
\caption{Statistics of the SMEsD.}
\resizebox{0.49\textwidth}{!}{
\begin{tabular}{cc|c|c|c}
\toprule
\multicolumn{2}{c|}{\textbf{SMEsD}}                                                & \textbf{Train} & \textbf{Valid} & \textbf{Test} \\ \midrule
\multicolumn{1}{c|}{\multirow{2}{*}{Node}}   
& \textit{}{company}            & 2,816  & 686   & 474  \\
\multicolumn{1}{c|}{}                        
& \textit{}{person}             & 1,752  & 346   & 316  \\ \midrule
\multicolumn{1}{c|}{\multirow{10}{*}{HeteG}}  
& \textit{}{manager}            & 2,658   & 694   & 540   \\
\multicolumn{1}{c|}{}
& \textit{}{shareholder}        & 4,002   & 944   & 666  \\
\multicolumn{1}{c|}{}                        
& \textit{}{other stakeholder}  & 4,426  & 976   & 908  \\
\multicolumn{1}{c|}{}                        
& \textit{}{holder\_investor\_company} & 1,440  & 366   & 130  \\
\multicolumn{1}{c|}{}                        
& \textit{}{holder\_investor\_person} & 5,186  & 1,118   & 1,010  \\
\multicolumn{1}{c|}{}                        
& \textit{}{branch}             & 594   & 88    & 74   \\
\multicolumn{1}{c|}{}                        
& \textit{}{loan}               & 346    & 82    & 52   \\
\multicolumn{1}{c|}{}                        
& \textit{}{deal}               & 46    & 4     & 18    \\
\multicolumn{1}{c|}{}                        
& \textit{}{rent}               & 8    & 0     & 8    \\
\multicolumn{1}{c|}{}                        
& \textit{}{creditor}           & 82    & 26    & 0   \\
 \midrule
\multicolumn{1}{c|}{\multirow{3}{*}{HyperG}} 
& \textit{}{industry}           & 108    & 82    & 68   \\
\multicolumn{1}{c|}{}                        
& \textit{}{area}               & 153   & 60    & 60   \\
\multicolumn{1}{c|}{}                        
& \textit{}{stakeholder}        & 756  & 157   & 148   \\ \midrule
\multicolumn{1}{c|}{\multirow{2}{*}{Label}}  
& \textit{}{bankrupt}           & 1,621  & 352   & 307  \\
\multicolumn{1}{c|}{}                        
& \textit{}{survive}            & 1,195  & 334   & 167  \\ \bottomrule
\end{tabular}
}
\label{tab: Statistics}
\end{table}

\textbf{SME: }In the shareholder networks, the number of edges between \textit{company\_individual\_company} (i.e., CICs edges) is 53,874, and the number of edges between \textit{company\_company} (i.e., CCs edges) is 15,908. In the board member and executives’ networks, the number of edges between \textit{company\_individual\_company} (i.e., CICb edges) is 139,413. Among these companies, 3,566 companies experienced bankruptcy. The dataset has been divided into training, validation, and test sets, with 9,181, 2,154, and 2,154 companies, respectively. 

\textbf{CSI300E: }The CSI300E dataset includes 185 listed companies and 275 corresponding executives. The main company relationships include Investment, Industry category, Supply chain, and Business partnership. The executive relationships primarily involve classmates and colleagues.

We summarize some popular financial network datasets in Table \ref{Dataset Analysis}. Due to the data privacy nature of financial scenarios, there are currently no larger publicly available datasets. Please note that our model leans towards global explanations and is not trained for individual explanatory instances. Therefore, the parameterized graph autoencoder and the feature masker within CF\textsuperscript{3} require only a minimal number of samples for effective training. These trained components can then be employed in an inductive setting to explain new instances. Importantly, the model parameter complexity of CF\textsuperscript{3} is independent of the input graph size. Consequently, the method we propose demonstrates good scalability.
\begin{table}[t]
\centering
\caption{Description of node characteristics.}
\label{tab: Description}
\resizebox{0.49\textwidth}{!}{
\begin{tabular}{cc|c}
\toprule
\multicolumn{2}{c|}{\textbf{Features}} & \textbf{Node dimension} \\ \midrule
\multicolumn{1}{c|}{\multirow{3}{*}{Business information}}   
& \textit{}{Registered capital}            & 0     \\
\multicolumn{1}{c|}{}                        
& \textit{}{Paid-in capital}             & 1     \\ 
\multicolumn{1}{c|}{}                        
& \textit{}{Established time}             & 2     \\ \midrule
\multicolumn{1}{c|}{\multirow{4}{*}{Litigation events}}  
& \textit{}{Lawsuit cause type}            & 3-13      \\
\multicolumn{1}{c|}{}                        
& \textit{}{Court level of lawsuit}        & 14-17      \\
\multicolumn{1}{c|}{}                        
& \textit{}{Verdict}  & 18-21     \\
\multicolumn{1}{c|}{}                        
& \textit{}{Duration of action} & 22    \\ \midrule
\multicolumn{1}{c|}{\multirow{3}{*}{Credit information}} 
& \textit{}{Trustworthiness incentives}           & 23-25     \\
\multicolumn{1}{c|}{}                        
& \textit{}{Administrative licenses}               & 26-30     \\
\multicolumn{1}{c|}{}                       
& \textit{}{Business exceptions}        & 31    \\ \bottomrule
\end{tabular}
}
\end{table}

\begin{table}[t]
  \centering
  \caption{Dataset Analysis.}
  \resizebox{0.49\textwidth}{!}{
    \begin{tabular}{c|c|c}
    \toprule
    \textbf{Literature} &  \textbf{Data}&  \textbf{\makecell[c]{Publicly\\Accessible}}  \\
    \midrule
   \cite{liang2016financial} & 239 bankrupt and 239 nonbankrupt cases& NO  \\
   \cite{veganzones2018investigation} &1,500 companies& NO  \\
   \cite{mai2019deep} & 11,827 firms and 94,994 firm-years& NO \\
   \cite{son2019data} & 977,940 companies& NO  \\
   \cite{yang2021financial} & 8.6 million SMEs& NO \\
    \cite{bi2022company} & 87,925 companies& NO  \\
    \cite{li2023learning} & More than 100 real-world industries& NO  \\
    \midrule
    CSI300E \cite{zhao2022stock}  (We use)&185 listed companies, 275 executives&  YES\\
    SME \cite{zheng2021heterogeneous} (We use)& 1,3489 companies      &  YES\\
     SMEsD \cite{wei2024combining}  (We use)&3,976 SMEs      &  YES\\
    \bottomrule
    \end{tabular}%
    }
  \label{Dataset Analysis}%
\end{table}


\subsection{Comparison with State of the Art}
\subsubsection{Baselines}To verify the effectiveness and superiority of the proposed method, we compare it with three groups of baselines: (1) factual inference methods, (2) counterfactual inference methods, and (3) large model inference methods. For factual inference, we select six state-of-the-art (SOTA) methods, including GNNExplainer \cite{ying2019gnnexplainer},  PGM-Explainer \cite{vu2020pgm}, PGExplainer \cite{luo2020parameterized}, Subgraphx \cite{yuan2021explainability}, Gem \cite{lin2021generative}, DnX \cite{pereira2023distill}, GOAt \cite{lugoat2023}, and GMT \cite{chen2024interpretable}. For counterfactual inference, we compare it with the current SOTA methods: CF-GNNExplainer \cite{lucic2022cf} and CF\textsuperscript{2} \cite{tan2022learning}. For inference methods based on large language models (LLMs), we select FinGPT \cite{yang2023fingpt}, SEP \cite{koa2024learning}, and GPT-4o \cite{openai2024api}. The detailed descriptions of the 13 SOTA baseline methods are as follows:

$\bullet$ \textbf{GNNExplainer} selects a compact sub-graph while maximizing the mutual information with the whole graph.

$\bullet$ \textbf{PGM-Explainer} can identify crucial graph components and generate an explanation in the form of a Probabilistic Graphical Model to approximate that prediction. 

$\bullet$ \textbf{PGExplainer} extends GNNExplainer by adopting a deep neural network to provide a global understanding of any GNN models on arbitrary machine learning tasks by collectively explaining multiple instances.

$\bullet$ \textbf{Subgraphx} uses Monte Carlo Tree Search
(MCTS) to find out the connected sub-graphs, and employs Shapley values \cite{kuhn1953contributions} to measure its importance by considering the interactions among different graph structures, which could preserve the predictions as explanations.

$\bullet$ \textbf{DnX} learns a surrogate GNN via knowledge distillation and extracts node or edge-level explanations by solving a simple convex program.

$\bullet$ \textbf{Gem} formulates the problem of explaining GNNs as a causal learning task. It then trains a causal explanation model equipped with a loss function based on Granger causality. Please note that our method differs from Gem in the following \textit{three aspects}: 1. Gem's attribution process is based on pairwise edges, whereas ours uses meta-paths; 2. Gem's graph generator works with homogeneous graphs, while ours is designed for heterogeneous graphs; 3. For evidence subgraphs, we combine counterfactual-factual reasoning with Granger causality, whereas Gem uses only Granger causality to extract subgraphs.

$\bullet$ \textbf{GOAt} computes the contribution of each node or edge feature to each scalar product and aggregates the contributions from all scalar products to derive the relative importance of each node and edge.

$\bullet$ \textbf{GMt} introduces a theoretical framework called SubMT, which formalizes interpretable subgraph learning through the multilinear extension of subgraph distribution, and proposes an XGNN architecture to better approximate SubMT.

$\bullet$ \textbf{CF-GNNExplainer} exclusively deletes edges within the input data to induce minimal perturbations, thereby altering the prediction outcome and subsequently generating counterfactual explanations.

$\bullet$ \textbf{CF\textsuperscript{2}} proposes a counterfactual and factual reasoning framework, which generates GNN explanations by simultaneously considering the necessity and sufficiency of the explanations.

$\bullet$ \textbf{FinGPT} is an instruction-tuned LLM model by ChatGLM, which can take in a series of market information to make risk predictions.

$\bullet$ \textbf{SEP} utilizes a language self-reflective agent and Proximal Policy Optimization (PPO) to enable the LLM to self-learn how to generate explainable stock predictions.

$\bullet$ \textbf{GPT-4o} is designed for natural human-computer interaction. It can accept multimodal inputs and generate multimodal outputs, serving as an enhanced version of GPT-4.
\subsubsection{Evaluation Metrics.}
Because both SMEsD and SME consist of real-world data without ground truth explanations, it is not suitable to use metrics such as accuracy and precision for evaluation. We use the $fid+$, $fid-$, charact, CEF, ROR, and Kappa score
to evaluate the performance of 
explanations in our experiments,
which are widely adopted in previous works  ~\cite{agarwal2023evaluating,li2023pen}.
The specific definitions of these metrics are as follows:

We follow previous work \cite{yuan2022explainability,amara2022graphframex,agarwal2023evaluating,li2023pen} and evaluate the performance of models by six metrics,  including  $fid+$\footnote{$fid+$ and $fid-$ are abbreviations for fidelity+ and fidelity-, respectively.}, $fid-$, charact, CEF, RoR, and Kappa. The detailed
definitions of six metrics are introduced below:
\begin{itemize}
\item[$\bullet$] The fidelity is to measure the impact of the generated interpretable subgraph on the initial predictions. $fid-$ represents prediction changes by keeping important input features and removing unimportant structures, and 
 $fid+$ indicates the difference in predicted probability between the original predictions and the new prediction after removing important input features.
\begin{equation}
\left\{
\begin{aligned}
fid+&=\frac{1}{N} \sum_{i=1}^N\left|\mathbb{I}\left(\hat{y}_i=y_i\right)-\mathbb{I}\left(\hat{y}_i^{G_{C \backslash S}}=y_i\right)\right| \\
fid-&=\frac{1}{N} \sum_{i=1}^N\left|\mathbb{I}\left(\hat{y}_i=y_i\right)-\mathbb{I}\left(\hat{y}_i^{G_S}=y_i\right)\right|\\
\end{aligned}
\right.
\end{equation}
\item[$\bullet$] Characterization score (charact) is a weighted harmonic mean of $fid+$ and $1-fid-$.
\begin{equation}
\textrm{ charact } = \frac{w_{+} + w_{-}}{\frac{w_{+}}{fid_{+}} + \frac{w_{-}}{1 - fid-}} \ ,
\end{equation}
where $w_+ = w_- = 0.5$.
\item[$\bullet$] GEF is utilized to quantify the unfaithfulness of the explanations to the underlying GNN predictor. A higher GEF score indicates a greater degree of unfaithfulness.
\begin{equation}
\textrm {GEF} = 1 - \exp^{(- KL(y || \hat{y}))} \ ,
\end{equation}
where $KL\left ( \ast\right )$ is the the KL-Divergence \cite{kullback1951information} function, which is used to measure the
similarity between two probability distributions.
\item[$\bullet$] RoR is the proportion of samples where the top risk source extracted by the explanation model matches at least one of the top picks of the three senior financial experts to the total number of samples.
\item[$\bullet$] Kappa score is used for consistency evaluation, calculating the average Fleiss’ kappa score between the recommendations by the algorithm and three seasoned financial experts.
\begin{equation}
    \kappa =\left (\bar{P}-\bar{P_{e}} \right )/ \left (1-\bar{P_{e}}  \right )
\end{equation}
where $1-\bar{P_{e}}$ gives the degree of
agreement that is attainable above chance, and $\bar{P}-\bar{P_{e}}$ gives the degree of agreement actually achieved above chance.
\end{itemize}
\subsection{Experimental Settings}
To ensure the reproducibility of the results, we provide detailed experimental settings. For fair comparison and computational efficiency, all explanation methods extract two-order subgraphs of the target nodes on both datasets. On the SMEsD dataset, we designate all test nodes as target company nodes to monitor sources of risk, while on the SME dataset, we randomly select 500 nodes from the test set as target company nodes. The number of edges in the interpretable computational subgraph on the SME dataset is configured to be 15 and that of SMEsD is 25. We set the edge type to 4 and the number of meta-paths to 3. The training epochs for both datasets are set to 200. The learning rate for the SMEsD dataset is set to 0.0015, and for the SME dataset, it is set to 0.002.

For FinGPT, SEP, and GPT-4o, we follow a three-step process for risk prediction and its interpretability analysis. First, we collect relational data of the target company, including the types of associations and the attributes of associated entities. Next, we use prompts to perform predictive analysis with the LLM, generating explanatory results. Finally, we have three financial experts to evaluate these explanatory results so as to derive evaluation metrics.
\subsection{RQ1: Comparison Experiment}
The comparison experiment results for all baselines and our proposed method on all three datasets are shown in \Cref{results_of_smesdp,results_of_hat,results_of_sme,CSI300E_of_dansmp}. the best results
are indicated in bold and the second-best results are underlined.
In summary, CF\textsuperscript{3} achieves the best results compared with the baselines on various datasets.

Specifically, compared to all other baseline models, our method performs optimally or equivalently on SMEsD. 
This is primarily attributed to the generated counterfactual graphs, simulating changes in the operational environment of companies, thereby enhancing the model performance.
Regarding metric $fid-$ and GEF, our model also achieves optimal or near-optimal performance, potentially due to our approach's inclination towards identifying risk information for company nodes from a counterfactual perspective. On SME, we observe similar experimental phenomena, but with a larger performance gain for our model. For instance, CF\textsuperscript{3} obtains a remarkable 39.71\% improvement in terms of $fid+$ and 50.00\% improvement in terms of $fid-$ compared to the best baseline. Additionally, all models exhibit comparatively lower performance on SME, possibly attributed to the inherent complexity of the dataset itself. Due to the strong language modeling and logical reasoning capabilities of large language models (LLMs), LLM-based models have demonstrated impressive explanation performance and consistency, with SEP achieving suboptimal performance. However, due to the complexity of financial data, LLMs are still unable to fully model the intricate interactions between multi-source heterogeneous data. Another observation is the outstanding performance of GNNExplainer, further emphasizing the significance of simultaneously considering node features and topological structure for company risk detection. Please note that we have not reported the results of SubgraphX on SME. This is because SubgraphX is designed specifically to handle a single graph and cannot directly perform masking operations on the SME dataset which contains three graphs. Additionally, since FinGPT, SEP, and GPT-4o cannot directly handle multi-source heterogeneous data, some of their results are not reported.

\subsection{RQ2: Ablation Studies}
To evaluate the effectiveness of different components in CF\textsuperscript{3}, we conduct ablation studies with the following variants: 1) \textbf{ W/O edge}  indicates that only layer-based feature mask are conducted; 2) \textbf{W/O feature} denotes considering only the heterogeneous edge; 3) \textbf{W/O attrib} does not involve the meta-path based attribution process;
4) \textbf{W/O fact} denotes the treatment with only counterfactual reasoning added; 5) \textbf{W/O counter} denotes the treatment with only factual reasoning added.

\begin{figure}[t]
  \centering
  \subfloat[Based on HAT.]
  {\includegraphics[width=0.235\textwidth]{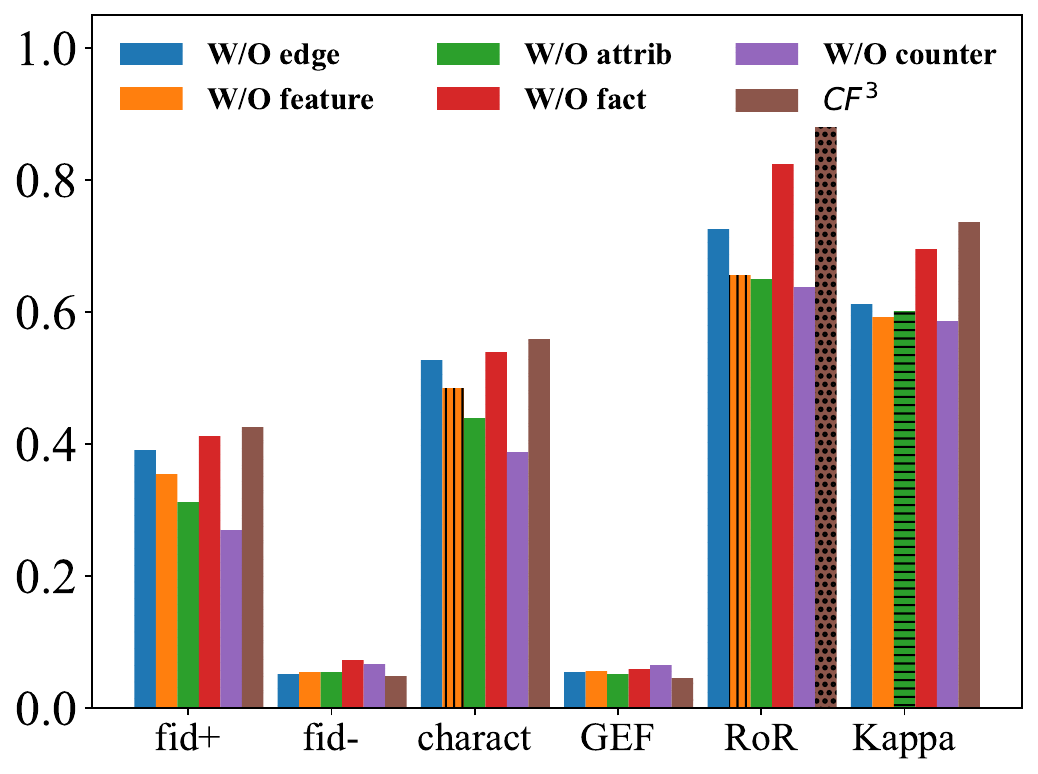}\label{ab-ahat}}
  \hfill    
  \subfloat[Based on ComRisk.]
  {\includegraphics[width=0.235\textwidth]{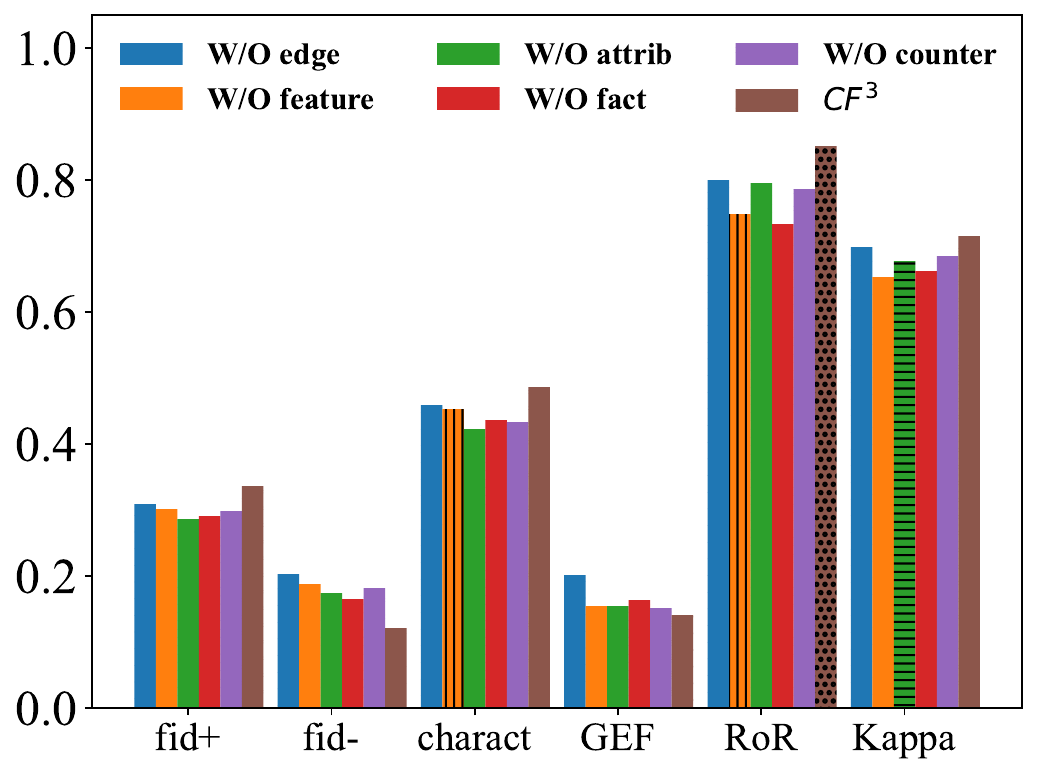}\label{ab-dcom}}
  \newline
  \subfloat[Based on HAT on SME.]
  {\includegraphics[width=0.235\textwidth]{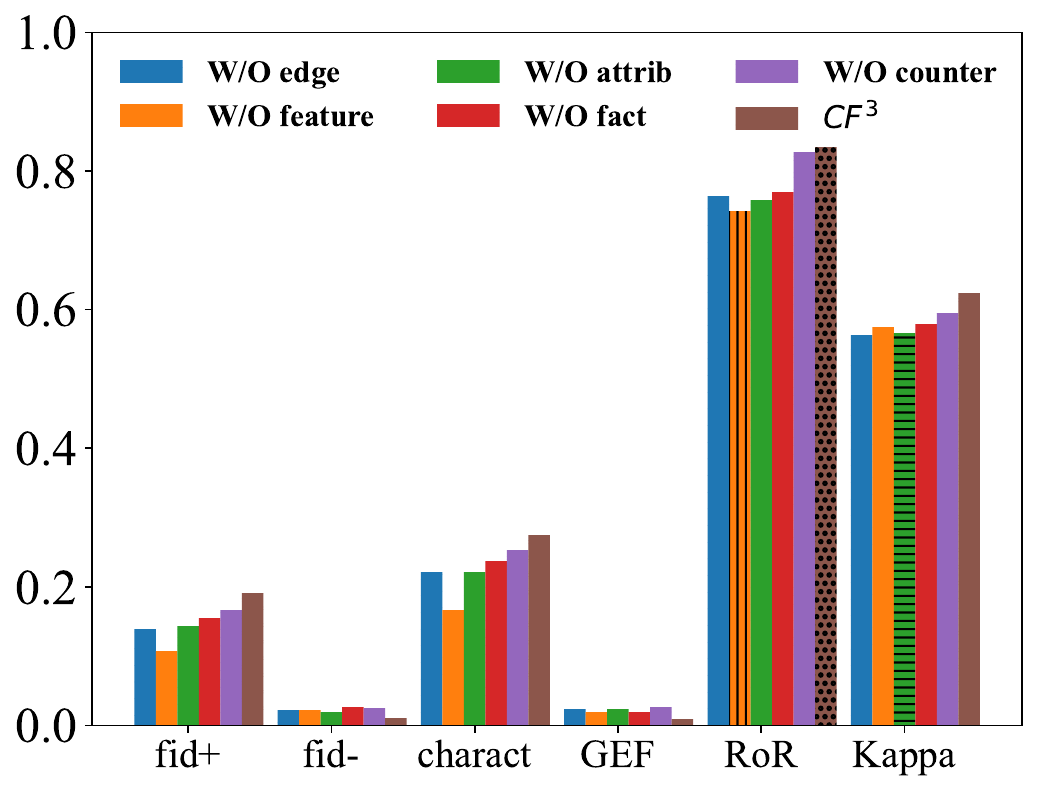}\label{ablation-shat}}
      \hfill
  \subfloat[Based on SME.]
  {\includegraphics[width=0.245\textwidth]{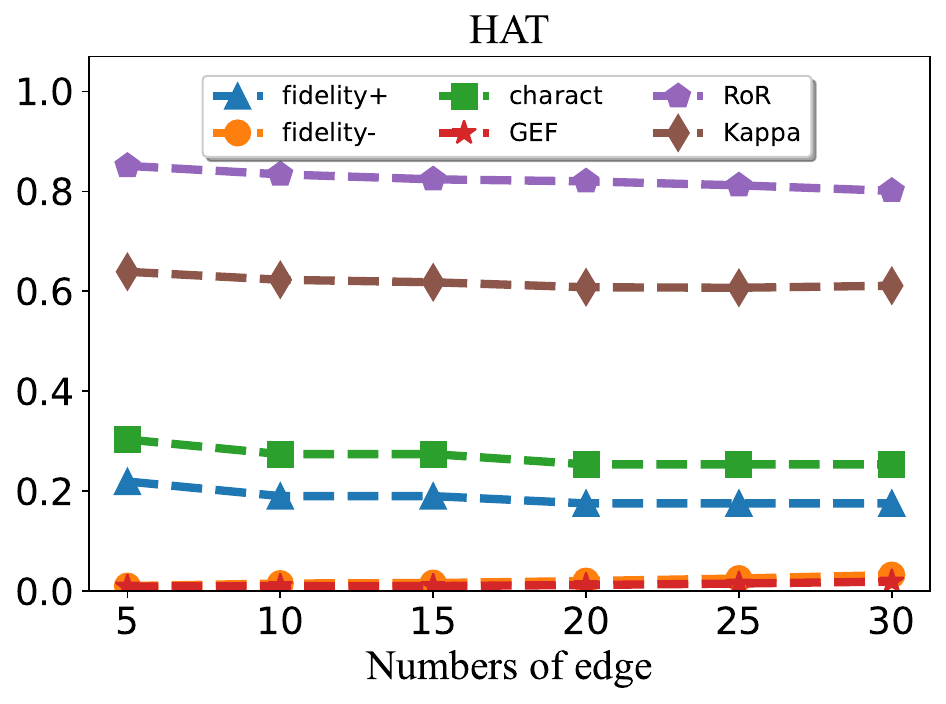}\label{SME-edge-figure}}
  \caption{(a)-(b) present the ablation analysis of CF\textsuperscript{3} on the SMEsD datasets, (c) on SME, and (d) shows the parameter analysis.}
  \label{Ablation analysis}
  \vspace{-1em}
\end{figure}

As depicted in  Figure \ref{ab-ahat} and \ref{ab-dcom}, we have the following observations: (1) No matter which part we remove in CF\textsuperscript{3}, the model’s performance drops. It indicates the effectiveness of all modules. (2) When we remove the meta-path based attribution module and counterfactual reasoning in CF\textsuperscript{3}, i.e., CF\textsuperscript{3} w/o attrib and
CF\textsuperscript{3} w/o counter, the performance drops more dramatically, indicating that the meta-path based attribution
process and counterfactual reasoning play a crucial role in financial risk detection.  (3) CF\textsuperscript{3} w/o edge and CF\textsuperscript{3} w/o feature are equally important. This implies that in the task of risk monitoring, significant node features and effective topological structures should be considered simultaneously. In addition, it can be seen from Figure \ref{ablation-shat} that in the SME dataset, the features of the company nodes have a greater impact on the model's predictions than the relationships between companies.
\begin{table}[t]
\caption{Performance comparison on SMEsD using ComRisk.}
\label{results_of_smesdp}

\resizebox{0.49\textwidth}{!}{
\begin{tabular}{l|cccc|cc}
 \toprule
\multicolumn{1}{l|}{\textbf{Explainer}} & \multicolumn{1}{c}{\textbf{$fid+$} $\uparrow$} & \multicolumn{1}{c}{\textbf{$fid-$} $\downarrow$} & \multicolumn{1}{c}{\textbf{charact} $\uparrow$} & \multicolumn{1}{c|}{\textbf{GEF} $\downarrow$}& \multicolumn{1}{c}{\textbf{ROR} $\uparrow$}& \multicolumn{1}{c}{\textbf{Kappa score
} $\uparrow$} \\ 
\midrule
\textbf{GNNExplainer}    & 0.263$\pm$0.005       & 0.133$\pm$0.005       & 0.407$\pm$0.008                  & 0.175$\pm$0.005 &0.784$\pm$0.015  &0.563$\pm$0.023    \\
\textbf{PGM-Explainer}    & 0.030$\pm$0.008       & 0.135$\pm$0.002       & 0.066$\pm$0.018                  & 0.168$\pm$0.001   & 0.631$\pm$0.011&  0.482$\pm$0.014  \\
\textbf{PGExplainer}      & 0.110$\pm$0.002      & 0.126$\pm$0.003       & 0.192$\pm$0.017                  & 0.229$\pm$0.005   &  0.607$\pm$0.008& 0.501$\pm$0.015   \\
\textbf{Subgraphx}        & 0.220$\pm$0.002       & 0.200$\pm$0.003       & 0.342$\pm$0.002                 & 0.350$\pm$0.002   &  0.725$\pm$0.022 &  0.511$\pm$0.018 \\
\textbf{DnX}              & 0.230$\pm$0.008       & \underline{0.123$\pm$0.005}      & 0.363$\pm$0.008                 & 0.181$\pm$0.009    & 0.761$\pm$0.012& 0.534$\pm$0.010  \\ 
\textbf{Gem}              & 0.279$\pm$0.009       & 0.125$\pm$0.006      & 0.418$\pm$0.008                 & \underline{0.157$\pm$0.007}&  0.627$\pm$0.014 &0.502$\pm$0.009    \\ 
\textbf{GOAt}             & \underline{0.311$\pm$0.017}       & 0.145$\pm$0.009      & 0.424$\pm$0.012                & 0.189$\pm$0.006&  0.805$\pm$0.013 &0.674$\pm$0.015\\
\textbf{GMT}             & \underline{0.313$\pm$0.011}       & 0.138$\pm$0.008      & \underline{0.441$\pm$0.009}                 & 0.175$\pm$0.006&  \underline{0.813$\pm$0.015} &\underline{0.686$\pm$0.012}\\
\midrule
 \textbf{CF-GNNExplainer} &0.026$\pm$0.008&0.167$\pm$0.009&0.045$\pm$0.012&0.221$\pm$0.009&0.624$\pm$0.026 &0.481$\pm$0.018\\
 \textbf{CF\textsuperscript{2}} &0.147$\pm$0.037&0.450$\pm$0.022&0.208$\pm$0.056&0.193$\pm$0.018&0.588$\pm$0.015&0.497$\pm$0.021 \\
\midrule
 \textbf{FinGPT} &-&-&-&-&0.621$\pm$0.027&0.504$\pm$0.020 \\
\textbf{SEP} &-&-&-&-&0.774$\pm$0.016& \underline{0.633$\pm$0.012}\\
  \textbf{GPT-4o} &-&-&-&-&0.745$\pm$0.019&0.568$\pm$0.014 \\
\midrule
\textbf{CF\textsuperscript{3} (ours)} & \textbf{0.336$\pm$0.012}       & \textbf{0.120$\pm$0.003}       & \textbf{0.486$\pm$0.009}                 & \textbf{0.140$\pm$0.015}&\textbf{0.851$\pm$0.017}&\textbf{0.715$\pm$0.025}
 \\
\bottomrule
\end{tabular}
}
\end{table}
\begin{table}[t]
\caption{
Experimental results on SMEsD using HAT.
}
\label{results_of_hat}
\resizebox{0.485\textwidth}{!}{
\begin{tabular}{l|cccc|cc}
\toprule
\multicolumn{1}{l|}{\textbf{Explainer}} & \multicolumn{1}{c}{\textbf{$fid+$} $\uparrow$} & \multicolumn{1}{c}{\textbf{$fid-$} $\downarrow$} & \multicolumn{1}{c}{\textbf{charact} $\uparrow$} & \multicolumn{1}{c|}{\textbf{GEF} $\downarrow$}& \multicolumn{1}{c}{\textbf{ROR} $\uparrow$}& \multicolumn{1}{c}{\textbf{Kappa score
} $\uparrow$} \\ 
\midrule
\textbf{GNNExplainer}     &0.340$\pm$0.016  & \textbf{0.043$\pm$0.005}    & 0.500$\pm$0.016               & 0.063$\pm$0.004  &  0.811$\pm$0.014& 0.664$\pm$0.009  \\
\textbf{PGM-Explainer}    & 0.077$\pm$0.012    & 0.063$\pm$0.005    & 0.147$\pm$0.005                 & 0.082$\pm$0.001  & 0.732$\pm$0.015 & 0.635$\pm$0.011   \\
\textbf{PGExplainer}      & 0.200$\pm$0.002    & 0.140$\pm$0.030    & 0.323$\pm$0.001                 & 0.125$\pm$0.004  & 0.746$\pm$0.013& 0.642$\pm$0.007   \\
\textbf{Subgraphx}        & 0.227$\pm$0.005    & 0.113$\pm$0.005    & 0.359$\pm$0.004                 & 0.115$\pm$0.003  & 0.751$\pm$0.015&  0.637$\pm$0.008   \\
\textbf{DnX}              & 0.287$\pm$0.009    & 0.053$\pm$0.005    & 0.443$\pm$0.007                & \underline{0.051$\pm$0.001}   & 0.762$\pm$0.014 &0.647$\pm$0.017  \\
\textbf{Gem}              & 0.363$\pm$0.009    & 0.064$\pm$0.005    & 0.521$\pm$0.007               & 0.073$\pm$0.001   & 0.768$\pm$0.016 & 0.653$\pm$0.010  \\
\textbf{GOAt}              &0.382$\pm$0.009    & 0.054$\pm$0.005    & 0.532$\pm$0.007                & 0.067$\pm$0.001   & 0.779$\pm$0.016 & 0.671$\pm$0.010  \\
\textbf{GMT}              & \underline{0.394$\pm$0.012}    & 0.059$\pm$0.013    & \underline{0.536$\pm$0.015}                & 0.081$\pm$0.011   & 0.816$\pm$0.019 & 0.691$\pm$0.016  \\
\midrule
 \textbf{CF-GNNExplainer} &0.101$\pm$0.010&0.063$\pm$0.014&0.105$\pm$0.007&0.108$\pm$0.005&0.782$\pm$0.009& 0.669$\pm$0.005\\
 \textbf{CF\textsuperscript{2}} &0.123$\pm$0.082&0.140$\pm$0.078&0.204$\pm$0.131&0.109$\pm$0.045 &0.793$\pm$0.010&0.682$\pm$0.007\\
 \midrule
\textbf{FinGPT} &-&-&-&-&0.794$\pm$0.022&0.691$\pm$0.018\\
\textbf{SEP} &-&-&-&-&\underline{0.827$\pm$0.012}&\underline{0.698$\pm$0.015}\\
 \textbf{GPT-4o} &-&-&-&-&0.786$\pm$0.014&0.674$\pm$0.017 \\
 \midrule
\textbf{CF\textsuperscript{3} (ours)} & \textbf{0.425$\pm$0.004}    & \underline{0.048$\pm$0.005}    & \textbf{0.559$\pm$0.006}                 & \textbf{0.044$\pm$0.008} &\textbf{0.879$\pm$0.021}&\textbf{0.736$\pm$0.028}\\
\bottomrule
\end{tabular}
}
\end{table}
\begin{table}[t]
\caption{
Performance comparisons on SME using HAT.}
\label{results_of_sme}
\resizebox{0.49\textwidth}{!}{
\begin{tabular}{l|cccc|cc}
 \toprule
\multicolumn{1}{l|}{\textbf{Explainer}} & \multicolumn{1}{c}{\textbf{$fid+$} $\uparrow$} & \multicolumn{1}{c}{\textbf{$fid-$} $\downarrow$} & \multicolumn{1}{c}{\textbf{charact} $\uparrow$} & \multicolumn{1}{c|}{\textbf{GEF} $\downarrow$} & \multicolumn{1}{c}{\textbf{ROR} $\uparrow$}& \multicolumn{1}{c}{\textbf{Kappa score
} $\uparrow$}\\ 
\midrule
\textbf{GNNExplainer}    & 0.080$\pm$0.008       & 0.033$\pm$0.005       & 0.150$\pm$0.012                  & \textbf{0.007$\pm$0.002} &   0.726$\pm$0.010 &  0.541$\pm$0.016  \\
\textbf{PGM-Explainer}    & 0.008$\pm$0.004       & 0.089$\pm$0.011      & 0.009$\pm$0.005                 & 0.054$\pm$0.009  & 0.617$\pm$0.021 &  0.495$\pm$0.011   \\
\textbf{PGExplainer}      & 0.013$\pm$0.005      & 0.023$\pm$0.005       & 0.025$\pm$0.007                  & 0.010$\pm$0.002  & 0.725$\pm$0.022 &  0.503$\pm$0.018   \\
\textbf{DnX}              & 0.020$\pm$0.003       & 0.030$\pm$0.008      & 0.039$\pm$0.004                 & 0.012$\pm$0.007 &   0.742$\pm$0.013&   0.511$\pm$0.017 \\ 
\textbf{Gem}              & 0.025$\pm$0.005       & 0.033$\pm$0.009      & 0.081$\pm$0.003                 & 0.031$\pm$0.004 &   0.624$\pm$0.009&   0.457$\pm$0.006 \\ 
\textbf{GOAt}              & 0.125$\pm$0.013       & 0.047$\pm$0.012      &0.194$\pm$0.009                & 0.043$\pm$0.009 &  0.795$\pm$0.013&   0.581$\pm$0.011 \\ 
\textbf{GMT}              & \underline{0.136$\pm$0.006}       & 0.038$\pm$0.007      & \underline{0.239$\pm$0.005}                 & 0.032$\pm$0.006 &   \underline{0.801$\pm$0.011}&   \underline{0.588$\pm$0.008} \\ 
\midrule
 \textbf{CF-GNNExplainer} &0.011$\pm$0.006&\underline{0.020$\pm$0.004}&0.025$\pm$0.012&0.014$\pm$0.006&0.653$\pm$0.010& 0.528$\pm$0.009\\
 \textbf{CF\textsuperscript{2}} &0.037$\pm$0.014&0.050$\pm$0.012&0.069$\pm$0.029&0.068$\pm$0.015& 0.761$\pm$0.008&0.547$\pm$0.015\\
 \midrule
   \textbf{FinGPT} &-&-&-&-&0.755$\pm$0.024&0.532$\pm$0.011 \\
\textbf{SEP} &-&-&-&-&0.774$\pm$0.024&0.561$\pm$0.013 \\
 \textbf{GPT-4o} &-&-&-&-&0.759$\pm$0.019&0.543$\pm$0.012 \\
 \midrule
\textbf{CF\textsuperscript{3} (ours)} & \textbf{0.190$\pm$0.011}       & \textbf{0.010$\pm$0.003}      & \textbf{0.274$\pm$0.013}                 & \underline{0.009$\pm$0.004}&\textbf{0.834$\pm$0.016}&\textbf{0.623$\pm$0.014}
 \\
\bottomrule
\end{tabular}
}
\end{table}
\begin{table}[t]
\caption{
Experimental results on CSI300E using DANSMP.
}
\label{CSI300E_of_dansmp}
\resizebox{0.49\textwidth}{!}{
\begin{tabular}{l|cccc|cc}
\toprule
\multicolumn{1}{l|}{\textbf{Explainer}} & \multicolumn{1}{c}{\textbf{$fid+$} $\uparrow$} & \multicolumn{1}{c}{\textbf{$fid-$} $\downarrow$} & \multicolumn{1}{c}{\textbf{charact} $\uparrow$} & \multicolumn{1}{c|}{\textbf{GEF} $\downarrow$}& \multicolumn{1}{c}{\textbf{ROR} $\uparrow$}& \multicolumn{1}{c}{\textbf{Kappa score
} $\uparrow$} \\ 
\midrule
\textbf{GNNExplainer}     &0.404$\pm$0.012  & \textbf{0.182$\pm$0.011}    & 0.534$\pm$0.011               & 0.158$\pm$0.015  &  0.819$\pm$0.010& 0.657$\pm$0.013  \\
\textbf{PGM-Explainer}    & 0.244$\pm$0.010    & 0.213$\pm$0.009    & 0.371$\pm$0.007                 & 0.197$\pm$0.011  & 0.724$\pm$0.018 & 0.621$\pm$0.015   \\
\textbf{PGExplainer}      & 0.347$\pm$0.014    & 0.206$\pm$0.024    & 0.478$\pm$0.011                 & 0.209$\pm$0.015  & 0.763$\pm$0.017& 0.647$\pm$0.007   \\
\textbf{Subgraphx}        & 0.318$\pm$0.008    & 0.194$\pm$0.009    & 0.451$\pm$0.011                 & 0.192$\pm$0.016  & 0.748$\pm$0.018&  0.631$\pm$0.013   \\
\textbf{DnX}              & 0.335$\pm$0.007    & 0.182$\pm$0.008    & 0.478$\pm$0.012                & \underline{0.147$\pm$0.011}   & 0.743$\pm$0.014 &0.640$\pm$0.014  \\
\textbf{Gem}              & 0.384$\pm$0.012    & 0.186$\pm$0.014    & 0.518$\pm$0.009               & 0.223$\pm$0.011   & 0.763$\pm$0.013 & 0.657$\pm$0.017  \\
\textbf{GOAt}              & 0.419$\pm$0.015   & 0.165$\pm$0.013    & 0.559$\pm$0.011                & 0.181$\pm$0.016   & 0.782$\pm$0.019 & 0.664$\pm$0.014  \\
\textbf{GMT}              & \underline{0.431$\pm$0.015}    & \underline{0.148$\pm$0.013}    & \underline{0.569$\pm$0.011}                & 0.153$\pm$0.016   & 0.811$\pm$0.019 & \underline{0.669$\pm$0.014}  \\
\midrule
 \textbf{CF-GNNExplainer} &0.252$\pm$0.011&0.194$\pm$0.010&0.387$\pm$0.009&0.212$\pm$0.011&0.770$\pm$0.012& 0.628$\pm$0.008\\
 \textbf{CF\textsuperscript{2}} &0.296$\pm$0.034&0.153$\pm$0.031&0.441$\pm$0.074&0.199$\pm$0.051 &0.781$\pm$0.011&0.653$\pm$0.017\\
 \midrule
\textbf{FinGPT} &-&-&-&-&\underline{0.819$\pm$0.019}&0.668$\pm$0.023\\
\textbf{SEP} &-&-&-&-&0.816$\pm$0.017&0.659$\pm$0.015\\
 \textbf{GPT-4o} &-&-&-&-&0.731$\pm$0.009&0.647$\pm$0.006 \\
 \midrule
\textbf{CF\textsuperscript{3} (ours)} & \textbf{0.466$\pm$0.008}    & \textbf{0.113$\pm$0.010}    & \textbf{0.602$\pm$0.009}                 & \textbf{0.087$\pm$0.011} &\textbf{0.841$\pm$0.013}&\textbf{0.696$\pm$0.019}\\
\bottomrule
\end{tabular}
}
\end{table}


\begin{figure}[t]
  \centering
  {\includegraphics[width=0.49\textwidth]{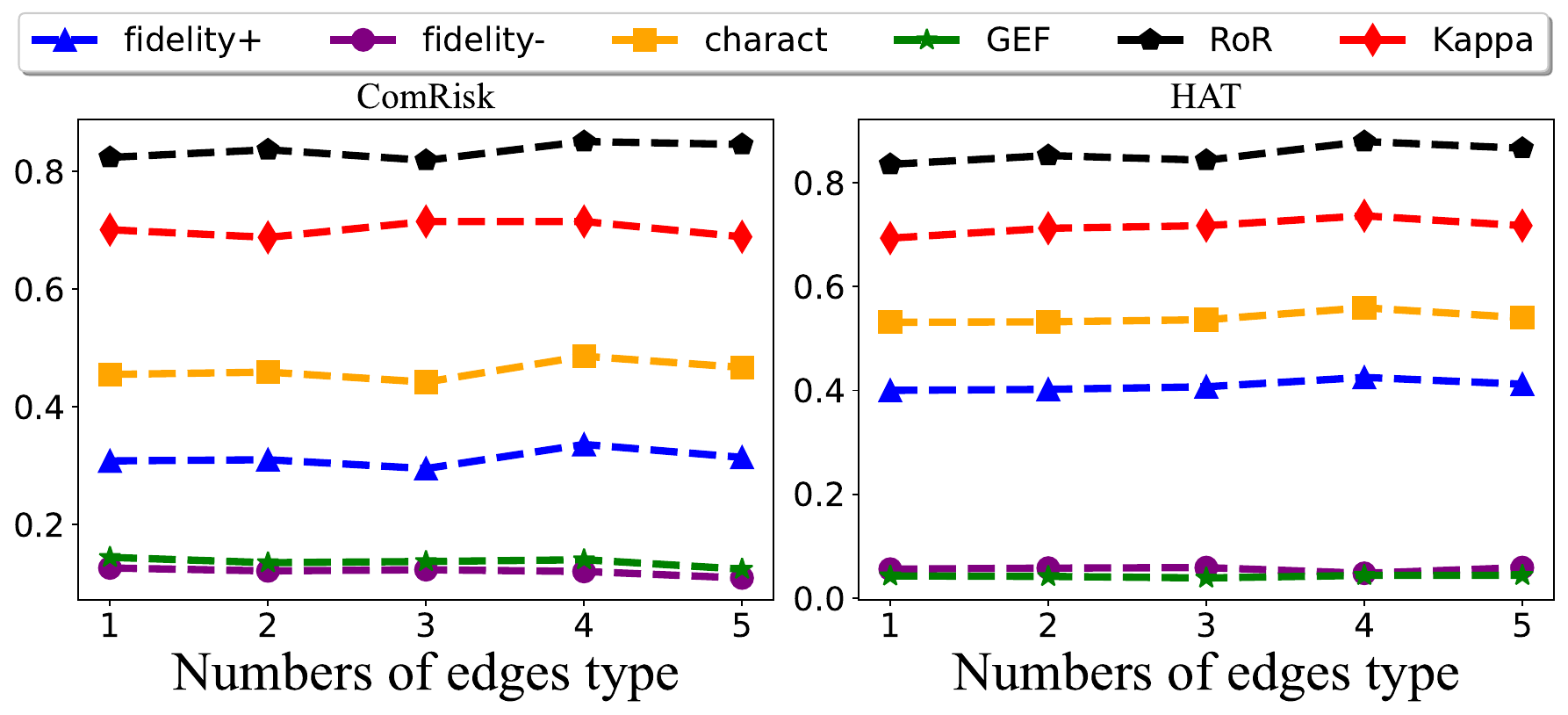}}
  \hfill    
  {\includegraphics[width=0.49\textwidth]{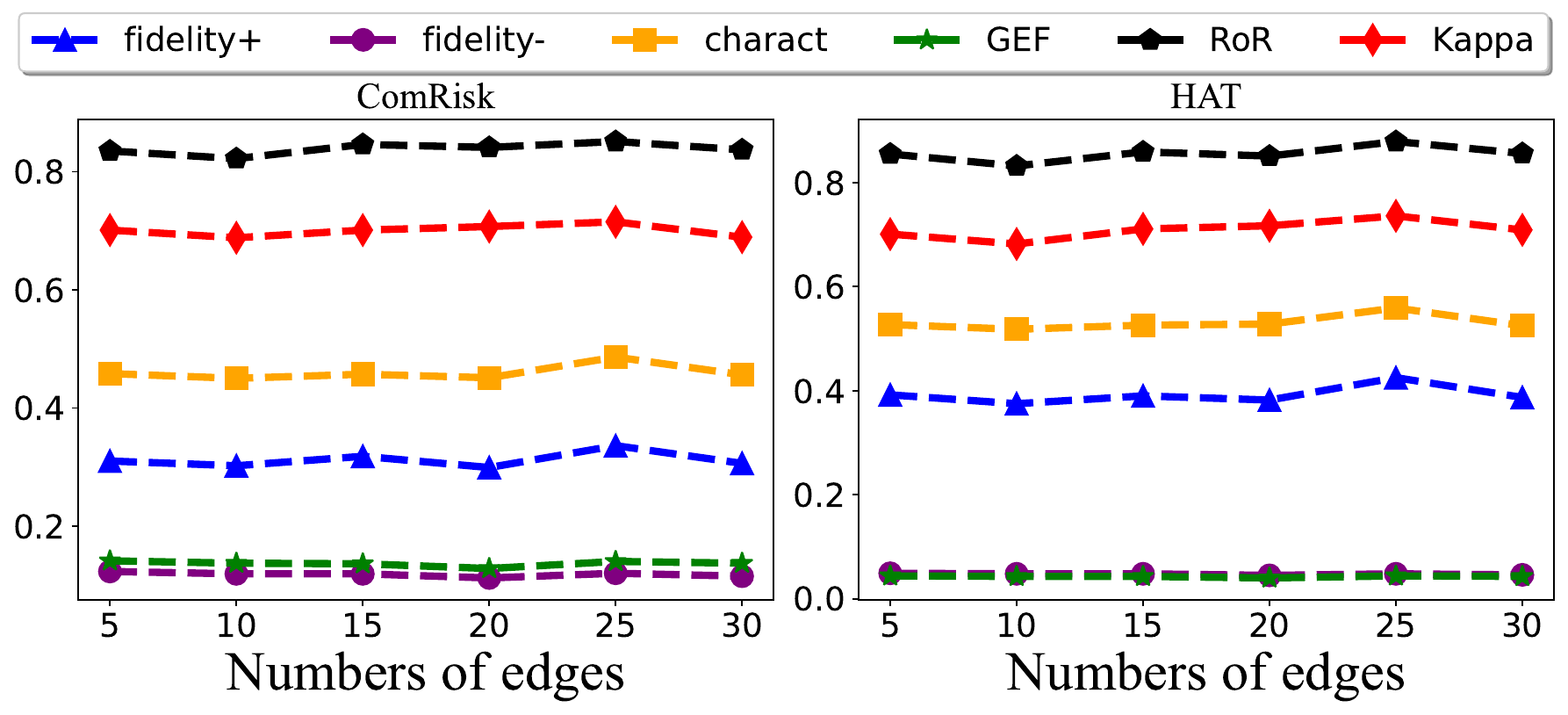}}
  \hfill
    {\includegraphics[width=0.49\textwidth]{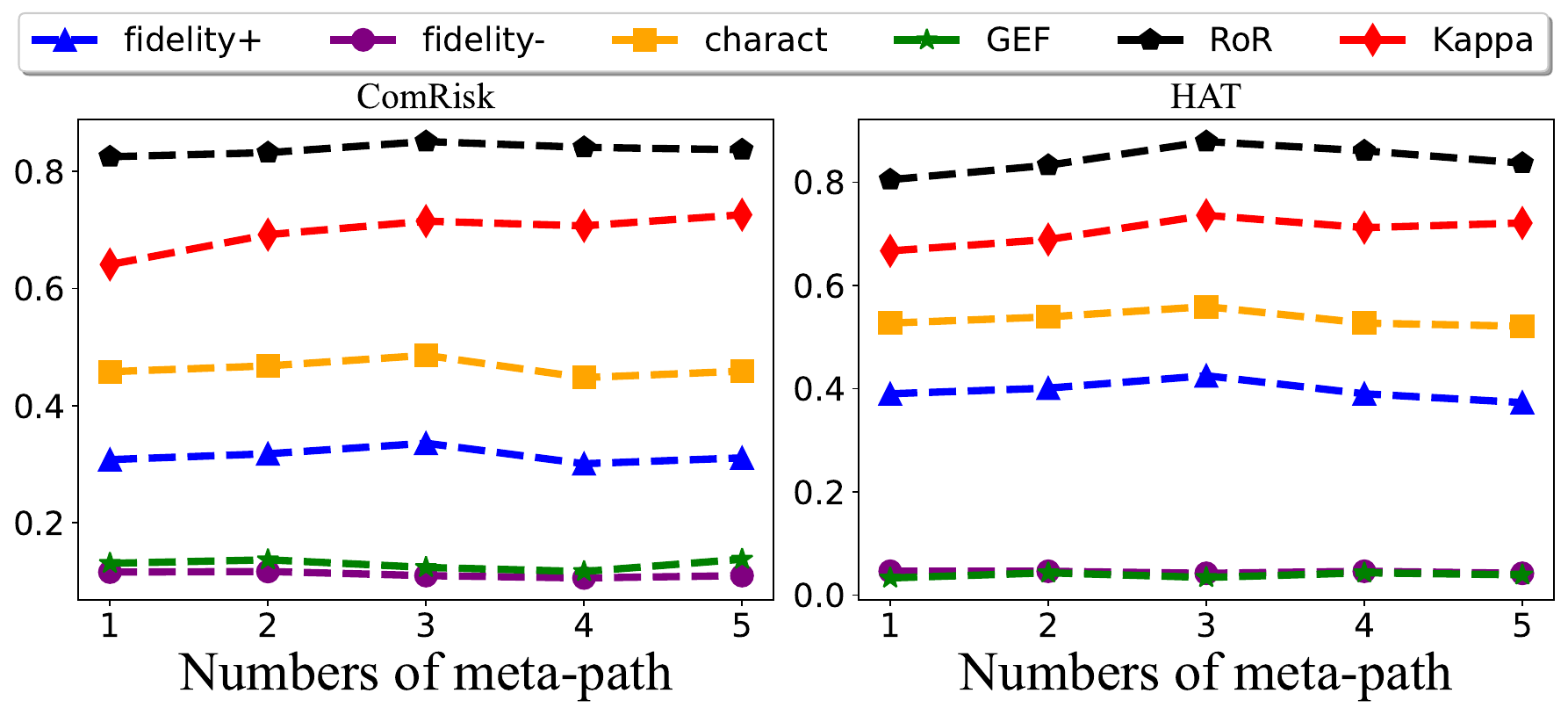}}
  \caption{The influence of different numbers of edge types and edges were selected for interpretation on SMEsD.}
  \label{parameter-smesd}
  \vspace{-1em}
\end{figure}
\subsection{RQ3: Parameter Analysis}
We also investigate the sensitivity analysis of edge types, the number of edges, and the number of meta-paths.
We report the results of CF\textsuperscript{3} under different parameter settings on both datasets. From the first row of Figure \ref{parameter-smesd}, it can be observed that the model performance initially improves and then decreases as the number of edge types increases. This trend may stem from the difficulty of overly simplified edge relationships in capturing the complexity of inter-company relationships. Additionally, we can infer that the performance of post-hoc explanation methods is influenced by the predictive model being explained. We did not report the ablation experiment results regarding the number of edge types on the SME dataset because this dataset lacks specific information about edge types.

In the second row of Figure \ref{parameter-smesd}, we can observed that as the number of edges increases, the performance of our model gradually improves. Notably, the variations in the values of $fid+$, charact, and Kappa are more pronounced, while the changes in the values of $fid-$, GEF, and RoR are relatively minor. This suggests that monitoring financial risk sources from a factual perspective requires identifying more business relationships. 

Here, we conducted further investigation into the impact of the number of meta-paths on the model performance. As depicted in Figure \ref{parameter-smesd}, it is evident that with an increase in the number of meta-paths, the model performance initially improves before declining. The model performance reaches its optimal level when the number of meta-paths reaches 3. However, in Figure \ref{SME-edge-figure}, the model performance decreases with an increase in the number of edges. This might be attributed to the significant noise in the SME dataset, where an excess of edges could introduce irrelevant noise and diminish the significance of the most crucial edges.

\subsection{RQ4: Computational performance}
CF\textsuperscript{3} and PGExplainer can explain previously unseen instances in the inductive setting. We measured the average inference time and training time for all methods, except for the LLMs-based methods that do not require training. As GNNExplainer explains one instance at a time, we measured its average explanation time cost for comparison. As reported in \ref{inference time}, it can be concluded that CF\textsuperscript{3} is able to provide more accurate explanations in a shorter amount of time.

The training time of all baselines are measured by reusing the source code released by the authors. Here, we take the SMEsD dataset as an example. \Cref{runtime-hat,runtime-comrisk} report the model's computational performance. We can conclude that CF\textsuperscript{3} amortizes the explanation cost by training a graph generator and layer-based feature masker to generate explanations for any given instances and consistently explain faster than the baseline overall. Please note that we have not reported the training times for FinGPT, SEP, and GPT-4o, as we utilized pre-trained versions of these models directly.

For CF\textsuperscript{3}, we measure the time for attributing the “ground-truth” explanation and the time for training the explainer with a varying number of training instances. As for GNNExplaner, we measure the overall time cost of explanations with a varying number of target instances. The training time of all baselines are measured by reusing the source code released by the authors. Here, we take the SMEsD dataset as an example. Tables \ref{runtime-hat} and \ref{runtime-comrisk} report the computational performance of all models. We can conclude that CF\textsuperscript{3} amortizes the explanation cost by training a graph generator and 
 layer-based feature masker to generate explanations for any given instances and consistently explain faster than the baseline overall. Please note that we have not reported the training times for FinGPT, SEP, and GPT-4o, as we utilized pre-trained versions of these models directly.
 \begin{table}[t]
  \centering
  \caption{Inference time per instance (s).}
  \resizebox{0.4\textwidth}{!}{
    \begin{tabular}{l|c|c}
    \toprule
    Datasets &  SMEsD (ComRisk)&  SMEsD (HAT)  \\
    \midrule
   \textbf{GNNExplainer} & 3.92& 8.18  \\
   \textbf{PGM-Explainer} & 4.85& 9.22  \\
   \textbf{PGExplainer} & \underline{1.06}& \underline{1.69}  \\
   \textbf{Subgraphx} & 3.63& 7.07  \\
   \textbf{Dnx} & 2.13& 2.85  \\
    \textbf{Gem} & 3.02& 5.74  \\
     \textbf{GOAt} & 7.38& 11.42  \\
     \textbf{GMt} & 10.24& 18.19  \\
    \textbf{CF-Explainer} & 180.32& 83.46  \\
   \textbf{CF\textsuperscript{2}} & 11.07& 14.33  \\
    \midrule
    \textbf{CF\textsuperscript{3} (ours)} & \textbf{0.24}      &  \textbf{0.46}\\
    \bottomrule
    \end{tabular}%
    }
  \label{inference time}%
\end{table}%
\begin{table}[t]
  \centering
  \caption{Training time ($\downarrow$) comparisons for all models on SMEsD based on HAT. Please note that the time units in the table are seconds}
  \resizebox{0.44\textwidth}{!}{
    \begin{tabular}{c|cccc}
    \toprule
    \makecell[c]{Number of \\training instances} & 100   & 200   & 300   & 400 \\
    \midrule
    \textbf{GNNExplainer} & 714.71 & 1636.84 & 2019.32 & 2941.65 \\
    \textbf{PGM-Explainer} & 542.78 & 1600.24 & 2378.81 & 2847.78 \\
    \textbf{PGExplainer} & 111.98 & 359.97 & 643.84 & 738.52 \\
    \textbf{Subgraphx} & 707.35 & 1043.28 & 4908.34 & 6820.14 \\
    \textbf{Dnx}   & 490.57 & 551.19 & 1104.14 & 1352.11 \\
    \textbf{Gem}   &171.43  & 361.24 & 502.19 & 714.16 \\
    \textbf{GAOt}   & 278.43 & 584.67 & 974.51 & 1161.56 \\
    \textbf{GMT}   &376.48  & 978.85 & 1204.74 & 1608.86 \\
    \textbf{CF-Explainer}    & 20316.70 & 40633.40 & 60950.11 & 81266.83 \\
    \textbf{CF\textsuperscript{2}}   & 1639.27 & 2865.11 & 14802.98 & 18674.87 \\
     \midrule
    \textbf{CF\textsuperscript{3} (ours)}   & \textbf{142.41} & \textbf{286.382} & \textbf{429.54} & \textbf{579.15} \\
    \bottomrule
    \end{tabular}
  }
    \label{runtime-hat}
\end{table}%
\begin{table}[t]
  \centering
  \caption{Training time ($\downarrow$) comparisons for all models on SMEsD based on  ComRisk. Please note that the time units in the table are seconds.}
  \resizebox{0.42\textwidth}{!}{
    \begin{tabular}{c|cccc}
    \toprule
    \makecell[c]{Number of \\training instances} & 100   & 200   & 300   & 400 \\
    \midrule
    \textbf{GNNExplainer} & 1710.52 & 3301.95 & 4914.17 & 6666.07 \\
    \textbf{PGM-Explainer} & 7702.76 & 15963.37 & 24245.83 & 31688.52 \\
    \textbf{PGExplainer} & 279.48 & 551.97 & 865.94 & 1125.82 \\
    \textbf{Subgraphx} & 5272.48 & 13962.85 & 22185.53 & 26198.78 \\
    \textbf{Dnx}   & 822.91 & 2925.54 & 5071.84 & 5301.71 \\
    \textbf{Gem}   & \underline{231.24} & \underline{496.27} & \underline{723.41} &  \underline{1014.32}\\
    \textbf{GOAt}   & 342.61 & 846.53 & 1130.61 &  1503.32\\
    \textbf{GMT}   & 411.14 & 915.08 & 1396.88 &  1685.17\\
    \textbf{CF-Explainer}    & 8273.87 & 16691.61 & 24929.52 & 33239.36 \\
    \textbf{CF\textsuperscript{2}}   & 1080.29 & 4050.33 & 4893.74 & 6315.58 \\
     \midrule
    \textbf{CF\textsuperscript{3} (ours)}   & \textbf{166.03} & \textbf{334.16} & \textbf{497.59} & \textbf{668.16} \\
    \bottomrule
    \end{tabular}}%
  \label{runtime-comrisk}%
\end{table}%
\begin{figure}[t]
	\centering
    \includegraphics[width=0.485\textwidth]{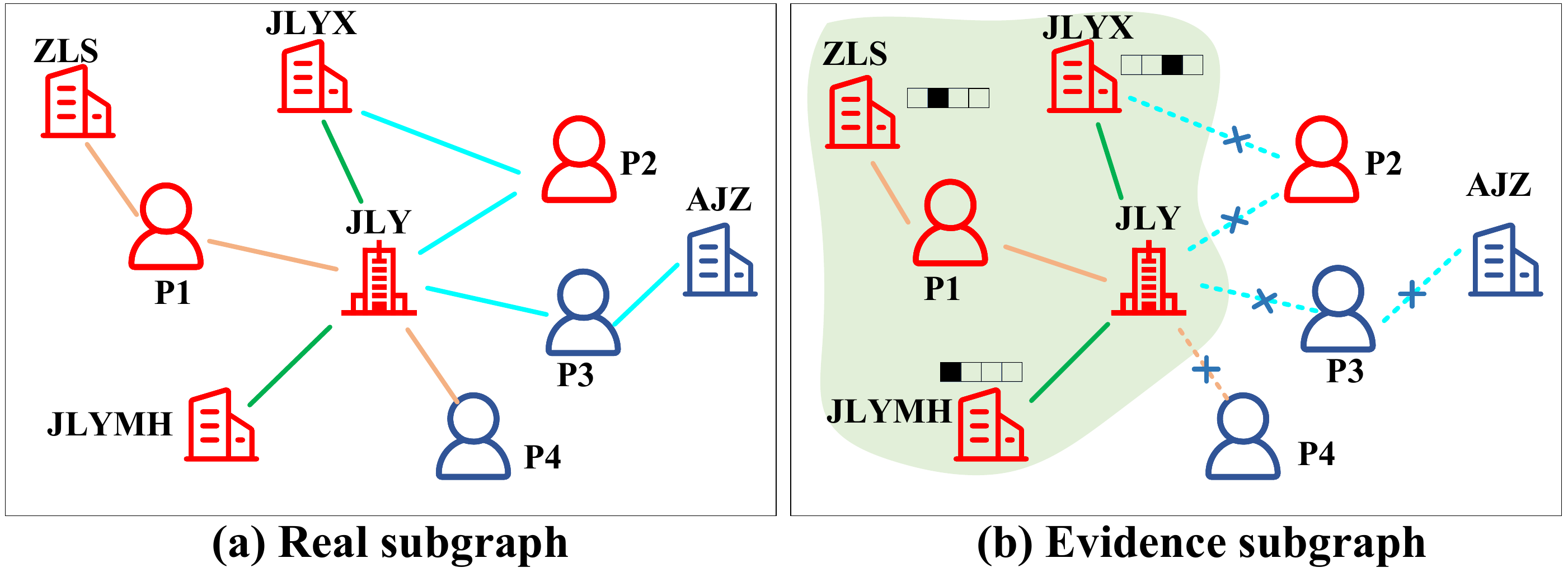}
	\caption{Structures of the real subgraphs and evidence subgraphs. And the dotted lines denote the removed edges.}
	\label{casestudy}
\end{figure}
\begin{table}[t]
  \centering
  \caption{The node features in raw and evidence subgraphs. In the table, "Raw" represents the raw subgraph, while "Extract" denotes the extracted evidence subgraph.}
  \resizebox{0.485\textwidth}{!}{
    \begin{tabular}{c|c|cccccc}
    \toprule
    & \textbf{State}    & \textbf{\makecell[c]{Company\\ node}}    & \textbf{\makecell[c]{Registered \\capital}}  & \textbf{\makecell[c]{Paid-in \\capital}}  & \textbf{\makecell[c]{Company \\litigation}}  &\textbf{\makecell[c]{Credit \\incentives}}  & \textbf{\makecell[c]{Administrative \\permits}}\\
    \midrule
    \multicolumn{1}{c|}{\multirow{4}[2]{*}{\makecell[c]{First-order\\ neighbors}}} & Raw     & JLYX     & 61    & 2     & 8    & 15    & 11 \\
          & Extract     & JLYX     & 61    & 0     & \cellcolor{blue!30}15     & \cellcolor{blue!30}0     & 11 \\
          & Raw     & JLYM     & 5     & 5     & 2     & 2     & 1 \\
          & Extract     & JLYM     & 5     & 5     & \cellcolor{blue!30}4     & \cellcolor{blue!30}0     & 1 \\
    \midrule
    \multicolumn{1}{c|}{\multirow{2}[2]{*}{\makecell[c]{Second-order \\neighbors}}} & Raw     & ZLS     & 10    & 10    & 1     & 1     & 1 \\
          & Extract     & ZLS     & 10    & 10     &1     & \cellcolor{blue!30}0     & \cellcolor{blue!30}0 \\
    \midrule
    \multicolumn{1}{c|}{\multirow{8}[2]{*}{\makecell[c]{The fused \\node features}}} & Raw     & JLY& 2     & 2     & 2     & 8     & 7 \\
          & Extract     & JLY & 2     & 2     & \cellcolor{green!30}9     & \cellcolor{green!30}0     & 7 \\
          & Raw     & JLYX     & 61    & 2     & 8    & 15    & 11 \\
          & Extract     & JLYX     & 61   & 2     & \cellcolor{green!30}10     & \cellcolor{green!30}0     & 11 \\
          & Raw     & JLYM     & 5     & 5     & 2     & 2     & 1 \\
          & Extract     & JLYM     & 5    & 5     & \cellcolor{green!30}6     & \cellcolor{green!30}0     & 1 \\
          & Raw     & ZLS     & 10    & 10    & 1     & 1     & 1 \\
          & Extract     & ZLS     & 10    & 10    & 0     & \cellcolor{green!30}0     & \cellcolor{green!30}0 \\
    \bottomrule
    \end{tabular}}
  \label{node feature}
\end{table}%
\subsection{RQ5: Case Study}
To validate the effectiveness of CF\textsuperscript{3}, we conducted a detailed study using JLY company from the SMEsD dataset as an example. In our illustration, red nodes represent risky entities, while blue nodes represent normal entities. In Figure \ref{casestudy}a, the two subsidiaries directly connected to JLY, JLYX and JLYMH, have declared bankruptcy. ZLS shares a common executive P1 with JLY. Figure \ref{casestudy}b displays the interpretable subgraph obtained by our method for the JLY company. We observe that the edges between JLY and JLYX, JLYMH, and ZLS have been retained. This is reasonable because if subsidiaries go bankrupt, the parent company is more likely to be affected by bankruptcy risk. These two companies have common executives, and if the executives are poorly managed, it will lead to risks for both companies. The removed portion in Figure \ref{casestudy}b does not provide useful information for predicting the bankruptcy of the JLY. For example, P2 holds a small amount of shares in both JLY and JLYX. Even if P2 is involved in some debt disputes, this will not affect JLY and JLYX through the shared ownership relationship. This observation indicates that CF\textsuperscript{3} can accurately identify factors related to the bankruptcy of the JLY and effectively eliminate irrelevant factors.

We further explore the decision-making process of the predictive model from the perspective of node features. These mainly include registered capital, paid-in capital, types of litigation, credit incentives, and administrative permits. In Figure \ref{casestudy}(a), company JLYX and JLYMH, which have already gone bankrupt, are two subsidiaries of Company JLY. And ZLS and LY share one executive P1. The original node features and the important node features extracted by our method are shown in Table \ref{node feature}. As shown in Table \ref{node feature}, our proposed explanatory method can identify the most relevant features influencing the labels of target company nodes, For instance, such as credit incentives and company litigation, which align with real-world scenarios. \textit{Intrinsic risk}: The target company exhibits a poor credit history and is involved in some company litigations, indicating potential bankruptcy risks. \textit{Contagion risk}: The associated companies of the target company share similar risk factors, which can propagate to the target company, further exacerbating its bankruptcy risk. By analyzing the features of different-order neighbors, we observe that for the target company, the credit incentives and  company litigation of first-order neighboring company nodes influence the target company. Meanwhile, the credit incentives and administrative permits of second-order neighboring company nodes significantly impact the target company's bankruptcy, thus further validating the effectiveness of our proposed layer-based feature masker. The explanations provided by our method can offer suggestions for preventing and mitigating company risks. If the company maintains normal production and operations and avoids lawsuits related to company debt, it will not face the risk of bankruptcy.
\subsection{Predictive Accuracy}
In Table \ref{predict performance}, we report the prediction results of the base model we used in this paper. It can be concluded that the prediction model we selected has good classification performance, which serves as the foundation for obtaining more accurate interpretative results. We use the same base model for all the baselines to fairly compare the risk detection ability. Since the explanation method is model-agnostic, the base model can be any classification model for node. To comprehensively demonstrate the effectiveness of our proposed method, we also conducted tests on various baseline models. Specifically, we applied the HAT and ComRisk models to the SMEsD dataset.
\begin{table}[htbp]
\caption{
The prediction result of HAT and ComRisk on both datasets.}
\label{predict performance}
\resizebox{0.49\textwidth}{!}{
\begin{tabular}{l|ccccccc}
 \toprule
\multicolumn{1}{l|}{\textbf{Models}} &
\multicolumn{1}{c}{\textbf{Datasets}} &
\multicolumn{1}{c}{\textbf{Accuracy}} & \multicolumn{1}{c}{\textbf{Precision}} & \multicolumn{1}{c}{\textbf{Recall}} & \multicolumn{1}{c}
{\textbf{F1}} & \multicolumn{1}{c}
{\textbf{AUC}} \\ 
 \midrule
\textbf{HAT}     &SMEsD  & 0.7785                       & 0.7701                        & 0.9381                     & 0.8458                 & 0.7960                  \\
\textbf{ComRisk} &SMEsD & 0.8397                       & 0.8348                        & 0.9381                     & 0.8834                 & 0.8862  \\
\textbf{HAT} &SME & 0.7794                       & 0.7447                        & 0.6178                     & 0.6315                 & 0.6178  \\
\textbf{DANSMP} &CSI300E & 0.6116                       & 0.7235                        & 0.6178                     & 0.6619                 & 0.6237  \\
\bottomrule
\end{tabular}
}
\end{table}
\section{Conclusion}
In this paper, we investigate company financial risk detection based on company knowledge graphs. First, adhering to the Granger causality principle, we employ a meta-path-based attribution process to select the most relevant meta-paths to construct an attribution subgraph, guiding the subsequent supervised learning process. Following this, we utilize an edge-type graph generator for graph reconstruction. For node features, we introduce a layer-based masking matrix, aiming to modify the node feature space. Finally, leveraging both counterfactual and factual reasoning, we extract evidence subgraphs containing significant subgraph structures and key node features for monitoring company financial risk. Extensive experiments on two public datasets demonstrate that our model provides better interpretations for company financial risk detection.
\bibliographystyle{IEEEtran}
\bibliography{TKDE-CF3}

\begin{thebibliography}{10}
\providecommand{\url}[1]{#1}
\csname url@samestyle\endcsname
\providecommand{\newblock}{\relax}
\providecommand{\bibinfo}[2]{#2}
\providecommand{\BIBentrySTDinterwordspacing}{\spaceskip=0pt\relax}
\providecommand{\BIBentryALTinterwordstretchfactor}{4}
\providecommand{\BIBentryALTinterwordspacing}{\spaceskip=\fontdimen2\font plus
\BIBentryALTinterwordstretchfactor\fontdimen3\font minus \fontdimen4\font\relax}
\providecommand{\BIBforeignlanguage}[2]{{%
\expandafter\ifx\csname l@#1\endcsname\relax
\typeout{** WARNING: IEEEtran.bst: No hyphenation pattern has been}%
\typeout{** loaded for the language `#1'. Using the pattern for}%
\typeout{** the default language instead.}%
\else
\language=\csname l@#1\endcsname
\fi
#2}}
\providecommand{\BIBdecl}{\relax}
\BIBdecl

\bibitem{liu2023qtiah}
Y.~Liu, Z.~Gao, X.~Liu, P.~Luo, Y.~Yang, and H.~Xiong, ``Qtiah-gnn: Quantity and topology imbalance-aware heterogeneous graph neural network for bankruptcy prediction,'' in \emph{Proceedings of the SIGKDD}, 2023, pp. 1572--1582.

\bibitem{kou2021bankruptcy}
G.~Kou, Y.~Xu, Y.~Peng, F.~Shen, Y.~Chen, K.~Chang, and S.~Kou, ``Bankruptcy prediction for smes using transactional data and two-stage multiobjective feature selection,'' \emph{Decision Support Systems}, vol. 140, p. 113429, 2021.

\bibitem{zhao2023stock}
Y.~Zhao, H.~Du, Y.~Liu, S.~Wei, X.~Chen, F.~Zhuang, Q.~Li, and G.~Kou, ``Stock movement prediction based on bi-typed hybrid-relational market knowledge graph via dual attention networks,'' \emph{IEEE Transactions on Knowledge and Data Engineering}, vol.~35, no.~8, pp. 8559--8571, 2023.

\bibitem{wei2024combining}
S.~Wei, J.~Lv, Y.~Guo, Q.~Yang, X.~Chen, Y.~Zhao, Q.~Li, F.~Zhuang, and G.~Kou, ``Combining intra-risk and contagion risk for enterprise bankruptcy prediction using graph neural networks,'' \emph{Information Sciences}, p. 120081, 2024.

\bibitem{zhang2022survey}
W.~Zhang, X.~Li, Y.~Deng, L.~Bing, and W.~Lam, ``A survey on aspect-based sentiment analysis: Tasks, methods, and challenges,'' \emph{IEEE Transactions on Knowledge and Data Engineering}, vol.~35, no.~11, pp. 11\,019--11\,038, 2023.

\bibitem{huang2023event}
H.~Huang, X.~Liu, G.~Shi, and Q.~Liu, ``Event extraction with dynamic prefix tuning and relevance retrieval,'' \emph{IEEE Transactions on Knowledge and Data Engineering}, vol.~35, no.~10, pp. 9946--9958, 2023.

\bibitem{xu2021rest}
W.~Xu, W.~Liu, C.~Xu, J.~Bian, J.~Yin, and T.-Y. Liu, ``Rest: Relational event-driven stock trend forecasting,'' in \emph{Proceedings of the Web Conference 2021}, 2021, pp. 1--10.

\bibitem{cheng2021modeling}
R.~Cheng and Q.~Li, ``Modeling the momentum spillover effect for stock prediction via attribute-driven graph attention networks,'' in \emph{Proceedings of the AAAI conference on artificial intelligence}, 2021, pp. 55--62.

\bibitem{bi2022company}
B.~Wendong, X.~Bingbing, S.~Xiaoqian, W.~Zidong, S.~Huawei, and C.~Xueqi, ``Company-as-tribe: Company financial risk assessment on tribe-style graph with hierarchical graph neural networks,'' in \emph{Proceedings of the 28th ACM SIGKDD Conference on Knowledge Discovery and Data Mining}, 2022, pp. 2712--2720.

\bibitem{wang2023financial}
D.~Wang, Z.~Zhang, Y.~Zhao, K.~Huang, Y.~Kang, and J.~Zhou, ``Financial default prediction via motif-preserving graph neural network with curriculum learning,'' in \emph{KDD}, 2023, pp. 2233--2242.

\bibitem{cho2023feature}
S.~H. Cho and K.-s. Shin, ``Feature-weighted counterfactual-based explanation for bankruptcy prediction,'' \emph{Expert Systems with Applications}, vol. 216, p. 119390, 2023.

\bibitem{nguyen2023bankruptcy}
H.~H. Nguyen, J.-L. Viviani, and S.~Ben~Jabeur, ``Bankruptcy prediction using machine learning and shapley additive explanations,'' \emph{Review of Quantitative Finance and Accounting}, pp. 1--42, 2023.

\bibitem{schnake2021higher}
T.~Schnake, O.~Eberle, J.~Lederer, S.~Nakajima, K.~T. Sch{\"u}tt, K.-R. M{\"u}ller, and G.~Montavon, ``Higher-order explanations of graph neural networks via relevant walks,'' \emph{IEEE transactions on pattern analysis and machine intelligence}, vol.~44, no.~11, pp. 7581--7596, 2021.

\bibitem{li2023pen}
S.~Li, W.~Liao, Y.~Chen, and R.~Yan, ``Pen: prediction-explanation network to forecast stock price movement with better explainability,'' in \emph{Proceedings of the AAAI Conference on Artificial Intelligence}, 2023, pp. 5187--5194.

\bibitem{koa2024learning}
K.~J. Koa, Y.~Ma, R.~Ng, and T.-S. Chua, ``Learning to generate explainable stock predictions using self-reflective large language models,'' in \emph{Proceedings of the ACM on Web Conference 2024}, 2024, pp. 4304--4315.

\bibitem{yu2024fusing}
G.~Yu, X.~Wang, Q.~Li, and Y.~Zhao, ``Fusing llms and kgs for formal causal reasoning behind financial risk contagion,'' \emph{arXiv preprint arXiv:2407.17190}, 2024.

\bibitem{wang2025risk}
J.~Wang, P.~Li, Y.~Liu, X.~Xiong, Y.~Zhang, and Z.~Lv, ``Risk identification of listed companies violation by integrating knowledge graph and multi-source risk factors,'' \emph{Engineering Applications of Artificial Intelligence}, vol. 141, p. 109774, 2025.

\bibitem{granger1980testing}
C.~W. Granger, ``Testing for causality: A personal viewpoint,'' \emph{Journal of Economic Dynamics and control}, vol.~2, pp. 329--352, 1980.

\bibitem{lin2021generative}
W.~Lin, H.~Lan, and B.~Li, ``Generative causal explanations for graph neural networks,'' in \emph{ICML}, 2021, pp. 6666--6679.

\bibitem{lueig2023}
S.~Lu, B.~Liu, K.~G. Mills, J.~He, and D.~Niu, ``Eig-search: Generating edge-induced subgraphs for gnn explanation in linear time,'' in \emph{Forty-first International Conference on Machine Learning}, 2023.

\bibitem{lugoat2023}
S.~Lu, K.~G. Mills, J.~He, B.~Liu, and D.~Niu, ``Goat: Explaining graph neural networks via graph output attribution,'' in \emph{The Twelfth International Conference on Learning Representations}, 2023.

\bibitem{azzolin2023global}
S.~Azzolin, A.~Longa, P.~Barbiero, P.~Lio, A.~Passerini \emph{et~al.}, ``Global explainability of gnns via logic combination of learned concepts,'' in \emph{ICLR}, 2023, pp. 1--19.

\bibitem{rong2024efficient}
Y.~Rong, G.~Wang, Q.~Feng, N.~Liu, Z.~Liu, E.~Kasneci, and X.~Hu, ``Efficient gnn explanation via learning removal-based attribution,'' \emph{ACM Transactions on Knowledge Discovery from Data}, 2024.

\bibitem{lu2024eig}
S.~Lu, B.~Liu, K.~G. Mills, J.~He, and D.~Niu, ``Eig-search: Generating edge-induced subgraphs for gnn explanation in linear time,'' in \emph{Forty-first International Conference on Machine Learning}, 2024.

\bibitem{armgaan2024graphtrail}
B.~Armgaan, M.~Dalmia, S.~Medya, and S.~Ranu, ``Graphtrail: Translating gnn predictions into human-interpretable logical rules,'' in \emph{The Thirty-eighth Annual Conference on Neural Information Processing Systems}, 2024.

\bibitem{chen2024interpretable}
Y.~Chen, Y.~Bian, B.~Han, and J.~Cheng, ``How interpretable are interpretable graph neural networks?'' in \emph{Forty-first International Conference on Machine Learning}, 2024.

\bibitem{tang2010market}
D.~Y. Tang and Y.~Hong, ``Market conditions, default risk and credit spreads,'' \emph{Journal of Banking \& Finance}, vol.~34, no.~4, pp. 743--753, 4 2010.

\bibitem{zheng2021heterogeneous}
Y.~Zheng, V.~C. Lee, Z.~Wu, and S.~Pan, ``Heterogeneous graph attention network for small and medium-sized enterprises bankruptcy prediction,'' in \emph{Pacific-Asia Conference on Knowledge Discovery and Data Mining}, 2021, pp. 140--151.

\bibitem{biddle2022accounting}
G.~C. Biddle, M.~L. Ma, and F.~M. Song, ``Accounting conservatism and bankruptcy risk,'' \emph{Journal of Accounting, Auditing \& Finance}, vol.~37, no.~2, pp. 295--323, 2022.

\bibitem{cheng2020contagious}
D.~Cheng, Z.~Niu, and Y.~Zhang, ``Contagious chain risk rating for networked-guarantee loans,'' in \emph{Proceedings of SIGKDD}, 2020, pp. 2715--2723.

\bibitem{Khashman2010NeuralNF}
A.~Khashman, ``Neural networks for credit risk evaluation: Investigation of different neural models and learning schemes,'' \emph{Expert Syst. Appl.}, vol.~37, pp. 6233--6239, 2010.

\bibitem{Cheng2022Regulating}
D.~Cheng, Z.~Niu, J.~Li, and C.~Jiang, ``Regulating systemic crises: Stemming the contagion risk in networked-loans through deep graph learning,'' \emph{IEEE Transactions on Knowledge and Data Engineering}, 2022.

\bibitem{Hu2020Loan}
B.~Hu, Z.~Zhang, J.~Zhou, J.~Fang, Q.~Jia, Y.~Fang, Q.~Yu, and Y.~Qi, ``Loan default analysis with multiplex graph learning,'' in \emph{Proceedings of CIKM}, 2020, pp. 2525--2532.

\bibitem{ye2020financial}
Y.~Zhen, Q.~Yu, and X.~Wei, ``Financial risk prediction with multi-round q\&a attention network.'' in \emph{Proceedings of IJCAI}, 2020, pp. 4576--4582.

\bibitem{wang2021ignorance}
Y.~Wang, J.~Li, D.~Wu, and R.~Anupindi, ``When ignorance is not bliss: An empirical analysis of subtier supply network structure on firm risk,'' \emph{Management Science}, vol.~67, no.~4, pp. 2029--2048, 2021.

\bibitem{gofman2022trade}
M.~Gofman and Y.~Wu, ``Trade credit and profitability in production networks,'' \emph{Journal of Financial Economics}, vol. 143, no.~1, pp. 593--618, 2022.

\bibitem{xu2021internet}
Y.~Xu, Y.~Xuan, and G.~Zheng, ``Internet searching and stock price crash risk: Evidence from a quasi-natural experiment,'' \emph{Journal of Financial Economics}, vol. 141, no.~1, pp. 255--275, 2021.

\bibitem{tobback2017bankruptcy}
E.~Tobback, T.~Bellotti, J.~Moeyersoms, M.~Stankova, and D.~Martens, ``Bankruptcy prediction for smes using relational data,'' \emph{Decision Support Systems}, vol. 102, pp. 69--81, 2017.

\bibitem{Yin2020Evaluating}
C.~Yin, C.~Jiang, H.~K. Jain, and Z.~Wang, ``Evaluating the credit risk of smes using legal judgments,'' \emph{Decis. Support Syst.}, vol. 136, p. 113364, 2020.

\bibitem{Yang2020Financial}
S.~Yang, Z.~Zhang, J.~Zhou, Y.~Wang, W.~Sun, X.~Zhong, Y.~Fang, Q.~Yu, and Y.~Qi, ``Financial risk analysis for smes with graph-based supply chain mining,'' in \emph{Proceedings of IJCAI}, 2021, pp. 4661--4667.

\bibitem{chuang2013application}
C.~Chun-Ling, ``Application of hybrid case-based reasoning for enhanced performance in bankruptcy prediction,'' \emph{Information Sciences}, vol. 236, pp. 174--185, 7 2013.

\bibitem{hu2018listening}
Z.~Hu, W.~Liu, J.~Bian, X.~Liu, and T.-Y. Liu, ``Listening to chaotic whispers: A deep learning framework for news-oriented stock trend prediction,'' in \emph{Proceedings of the eleventh ACM international conference on web search and data mining}, 2018, pp. 261--269.

\bibitem{dang2021squawk}
X.-H. Dang, S.~Y. Shah, and P.~Zerfos, ``" the squawk bot" joint learning of time series and text data modalities for automated financial information filtering,'' in \emph{Proceedings of the Twenty-Ninth International Conference on International Joint Conferences on Artificial Intelligence}, 2021, pp. 4597--4603.

\bibitem{yuan2021explainability}
H.~Yuan, H.~Yu, J.~Wang, K.~Li, and S.~Ji, ``On explainability of graph neural networks via subgraph explorations,'' in \emph{International Conference on Machine Learning}.\hskip 1em plus 0.5em minus 0.4em\relax PMLR, 2021, pp. 12\,241--12\,252.

\bibitem{huang2022graphlime}
Q.~Huang, M.~Yamada, Y.~Tian, D.~Singh, and Y.~Chang, ``Graphlime: Local interpretable model explanations for graph neural networks,'' \emph{IEEE Transactions on Knowledge and Data Engineering}, vol.~35, no.~7, pp. 6968--6972, 2022.

\bibitem{zhang2022protgnn}
Z.~Zhang, Q.~Liu, H.~Wang, C.~Lu, and C.~Lee, ``Protgnn: Towards self-explaining graph neural networks,'' in \emph{Proceedings of the AAAI Conference on Artificial Intelligence}, 2022, pp. 9127--9135.

\bibitem{zhang2022gstarx}
S.~Zhang, Y.~Liu, N.~Shah, and Y.~Sun, ``Gstarx: Explaining graph neural networks with structure-aware cooperative games,'' \emph{Advances in Neural Information Processing Systems}, vol.~35, pp. 19\,810--19\,823, 2022.

\bibitem{abrate2021counterfactual}
C.~Abrate and F.~Bonchi, ``Counterfactual graphs for explainable classification of brain networks,'' in \emph{KDD}, 2021, pp. 2495--2504.

\bibitem{bajaj2021robust}
M.~Bajaj, L.~Chu, Z.~Y. Xue, J.~Pei, L.~Wang, P.~C.-H. Lam, and Y.~Zhang, ``Robust counterfactual explanations on graph neural networks,'' in \emph{Advances in Neural Information Processing Systems}, 2021, pp. 5644--5655.

\bibitem{lucic2022cf}
A.~Lucic, M.~A. Ter~Hoeve, G.~Tolomei, M.~De~Rijke, and F.~Silvestri, ``Cf-gnnexplainer: Counterfactual explanations for graph neural networks,'' in \emph{International Conference on Artificial Intelligence and Statistics}, 2022, pp. 4499--4511.

\bibitem{ying2019gnnexplainer}
Z.~Ying, D.~Bourgeois, J.~You, M.~Zitnik, and J.~Leskovec, ``Gnnexplainer: Generating explanations for graph neural networks,'' \emph{Advances in neural information processing systems}, vol.~32, 2019.

\bibitem{luo2020parameterized}
D.~Luo, W.~Cheng, D.~Xu, W.~Yu, B.~Zong, H.~Chen, and X.~Zhang, ``Parameterized explainer for graph neural network,'' in \emph{Advances in neural information processing systems}, 2020, pp. 19\,620--19\,631.

\bibitem{miao2022interpretable}
S.~Miao, M.~Liu, and P.~Li, ``Interpretable and generalizable graph learning via stochastic attention mechanism,'' in \emph{International Conference on Machine Learning}.\hskip 1em plus 0.5em minus 0.4em\relax PMLR, 2022, pp. 15\,524--15\,543.

\bibitem{miao2023interpretable}
S.~Miao, Y.~Luo, M.~Liu, and P.~Li, ``Interpretable geometric deep learning via learnable randomness injection,'' in \emph{International Conference on Learning Representations}, 2023.

\bibitem{wu2022discovering}
Y.~Wu, X.~Wang, A.~Zhang, X.~He, and T.-S. Chua, ``Discovering invariant rationales for graph neural networks,'' in \emph{International Conference on Learning Representations}, 2022.

\bibitem{li2020survey}
X.-H. Li, C.~C. Cao, Y.~Shi, W.~Bai, H.~Gao, L.~Qiu, C.~Wang, Y.~Gao, S.~Zhang, X.~Xue \emph{et~al.}, ``A survey of data-driven and knowledge-aware explainable ai,'' \emph{IEEE Transactions on Knowledge and Data Engineering}, vol.~34, no.~1, pp. 29--49, 2020.

\bibitem{tan2022learning}
J.~Tan, S.~Geng, Z.~Fu, Y.~Ge, S.~Xu, Y.~Li, and Y.~Zhang, ``Learning and evaluating graph neural network explanations based on counterfactual and factual reasoning,'' in \emph{Proceedings of the ACM Web Conference}, 2022, pp. 1018--1027.

\bibitem{huang2023global}
Z.~Huang, M.~Kosan, S.~Medya, S.~Ranu, and A.~Singh, ``Global counterfactual explainer for graph neural networks,'' in \emph{Proceedings of the Sixteenth ACM International Conference on Web Search and Data Mining}, 2023, pp. 141--149.

\bibitem{zhao2022stock}
Y.~Zhao, H.~Du, Y.~Liu, S.~Wei, X.~Chen, F.~Zhuang, Q.~Li, and G.~Kou, ``Stock movement prediction based on bi-typed hybrid-relational market knowledge graph via dual attention networks,'' \emph{IEEE Transactions on Knowledge and Data Engineering}, vol.~35, no.~8, pp. 8559--8571, 2022.

\bibitem{wang2019heterogeneous}
X.~Wang, H.~Ji, C.~Shi, B.~Wang, Y.~Ye, P.~Cui, and P.~S. Yu, ``Heterogeneous graph attention network,'' in \emph{The world wide web conference}, 2019, pp. 2022--2032.

\bibitem{zhang2023page}
S.~Zhang, J.~Zhang, X.~Song, S.~Adeshina, D.~Zheng, C.~Faloutsos, and Y.~Sun, ``Page-link: Path-based graph neural network explanation for heterogeneous link prediction,'' in \emph{Proceedings of the ACM Web Conference 2023}, 2023, pp. 3784--3793.

\bibitem{goldstein2015peeking}
A.~Goldstein, A.~Kapelner, J.~Bleich, and E.~Pitkin, ``Peeking inside the black box: Visualizing statistical learning with plots of individual conditional expectation,'' \emph{journal of Computational and Graphical Statistics}, vol.~24, no.~1, pp. 44--65, 2015.

\bibitem{wang2021towards}
X.~Wang, Y.~Wu, A.~Zhang, X.~He, and T.-S. Chua, ``Towards multi-grained explainability for graph neural networks,'' in \emph{Advances in Neural Information Processing Systems}, 2021, pp. 18\,446--18\,458.

\bibitem{yang2023counterfactual}
Q.~Yang, C.~Ma, Q.~Zhang, X.~Gao, C.~Zhang, and X.~Zhang, ``Counterfactual learning on heterogeneous graphs with greedy perturbation,'' in \emph{Proceedings of the SIGKDD}, 2023, pp. 2988--2998.

\bibitem{kipf2016variational}
T.~N. Kipf and M.~Welling, ``Variational graph auto-encoders,'' in \emph{In NeurIPS Workshops}, 2016.

\bibitem{xiao2023counterfactual}
C.~Xiao, X.~Xu, Y.~Lei, K.~Zhang, S.~Liu, and F.~Zhou, ``Counterfactual graph learning for anomaly detection on attributed networks,'' \emph{IEEE Transactions on Knowledge and Data Engineering}, vol.~35, no.~8, pp. 8559--8571, 2023.

\bibitem{ma2022clear}
J.~Ma, R.~Guo, S.~Mishra, A.~Zhang, and J.~Li, ``Clear: Generative counterfactual explanations on graphs,'' \emph{Advances in Neural Information Processing Systems}, vol.~35, pp. 25\,895--25\,907, 2022.

\bibitem{zhang2023mixupexplainer}
J.~Zhang, D.~Luo, and H.~Wei, ``Mixupexplainer: Generalizing explanations for graph neural networks with data augmentation,'' in \emph{Proceedings of the SIGKDD}, 2023, pp. 3286--3296.

\bibitem{liang2016financial}
D.~Liang, C.-C. Lu, C.-F. Tsai, and G.-A. Shih, ``Financial ratios and corporate governance indicators in bankruptcy prediction: A comprehensive study,'' \emph{European journal of operational research}, vol. 252, no.~2, pp. 561--572, 2016.

\bibitem{veganzones2018investigation}
D.~Veganzones and E.~S{\'e}verin, ``An investigation of bankruptcy prediction in imbalanced datasets,'' \emph{Decision Support Systems}, vol. 112, pp. 111--124, 2018.

\bibitem{mai2019deep}
F.~Mai, S.~Tian, C.~Lee, and L.~Ma, ``Deep learning models for bankruptcy prediction using textual disclosures,'' \emph{European journal of operational research}, vol. 274, no.~2, pp. 743--758, 2019.

\bibitem{son2019data}
H.~Son, C.~Hyun, D.~Phan, and H.~J. Hwang, ``Data analytic approach for bankruptcy prediction,'' \emph{Expert Systems with Applications}, vol. 138, p. 112816, 2019.

\bibitem{yang2021financial}
S.~Yang, Z.~Zhang, J.~Zhou, Y.~Wang, W.~Sun, X.~Zhong, Y.~Fang, Q.~Yu, and Y.~Qi, ``Financial risk analysis for smes with graph-based supply chain mining,'' in \emph{Proceedings of the Twenty-Ninth International Conference on International Joint Conferences on Artificial Intelligence}, 2021, pp. 4661--4667.

\bibitem{li2023learning}
Y.~Li, Z.~Zhu, X.~Guo, L.~Chen, Z.~Wang, Y.~Wang, B.~Han, and Y.~Zhao, ``Learning joint relational co-evolution in spatial-temporal knowledge graph for smes supply chain prediction,'' in \emph{Proceedings of the 29th ACM SIGKDD Conference on Knowledge Discovery and Data Mining}, 2023, pp. 4426--4436.

\bibitem{vu2020pgm}
M.~Vu and M.~T. Thai, ``Pgm-explainer: Probabilistic graphical model explanations for graph neural networks,'' in \emph{Advances in neural information processing systems}, 2020, pp. 12\,225--12\,235.

\bibitem{pereira2023distill}
T.~Pereira, E.~Nascimento, L.~E. Resck, D.~Mesquita, and A.~Souza, ``Distill n’explain: explaining graph neural networks using simple surrogates,'' in \emph{International Conference on Artificial Intelligence and Statistics}, 2023, pp. 6199--6214.

\bibitem{yang2023fingpt}
H.~Yang, X.-Y. Liu, and C.~D. Wang, ``Fingpt: Open-source financial large language models,'' \emph{arXiv preprint arXiv:2306.06031}, 2023.

\bibitem{openai2024api}
{OpenAI}, ``{{OpenAI API}} models,'' 2024.

\bibitem{kuhn1953contributions}
H.~W. Kuhn and A.~W. Tucker, \emph{Contributions to the Theory of Games}.\hskip 1em plus 0.5em minus 0.4em\relax Princeton University Press, 1953, no.~28.

\bibitem{agarwal2023evaluating}
C.~Agarwal, O.~Queen, H.~Lakkaraju, and M.~Zitnik, ``Evaluating explainability for graph neural networks,'' \emph{Scientific Data}, vol.~10, no.~1, p. 144, 2023.

\bibitem{yuan2022explainability}
H.~Yuan, H.~Yu, S.~Gui, and S.~Ji, ``Explainability in graph neural networks: A taxonomic survey,'' \emph{IEEE transactions on pattern analysis and machine intelligence}, vol.~45, no.~5, pp. 5782--5799, 2022.

\bibitem{amara2022graphframex}
K.~Amara, R.~Ying, Z.~Zhang, Z.~Han, Y.~Shan, U.~Brandes, S.~Schemm, and C.~Zhang, ``Graphframex: Towards systematic evaluation of explainability methods for graph neural networks,'' \emph{arXiv preprint arXiv:2206.09677}, 2022.

\bibitem{kullback1951information}
S.~Kullback and R.~A. Leibler, ``On information and sufficiency,'' \emph{The annals of mathematical statistics}, vol.~22, no.~1, pp. 79--86, 1951.

\end{thebibliography}

\begin{IEEEbiography}[{\includegraphics[width=1in,height=1.25in,clip,keepaspectratio]{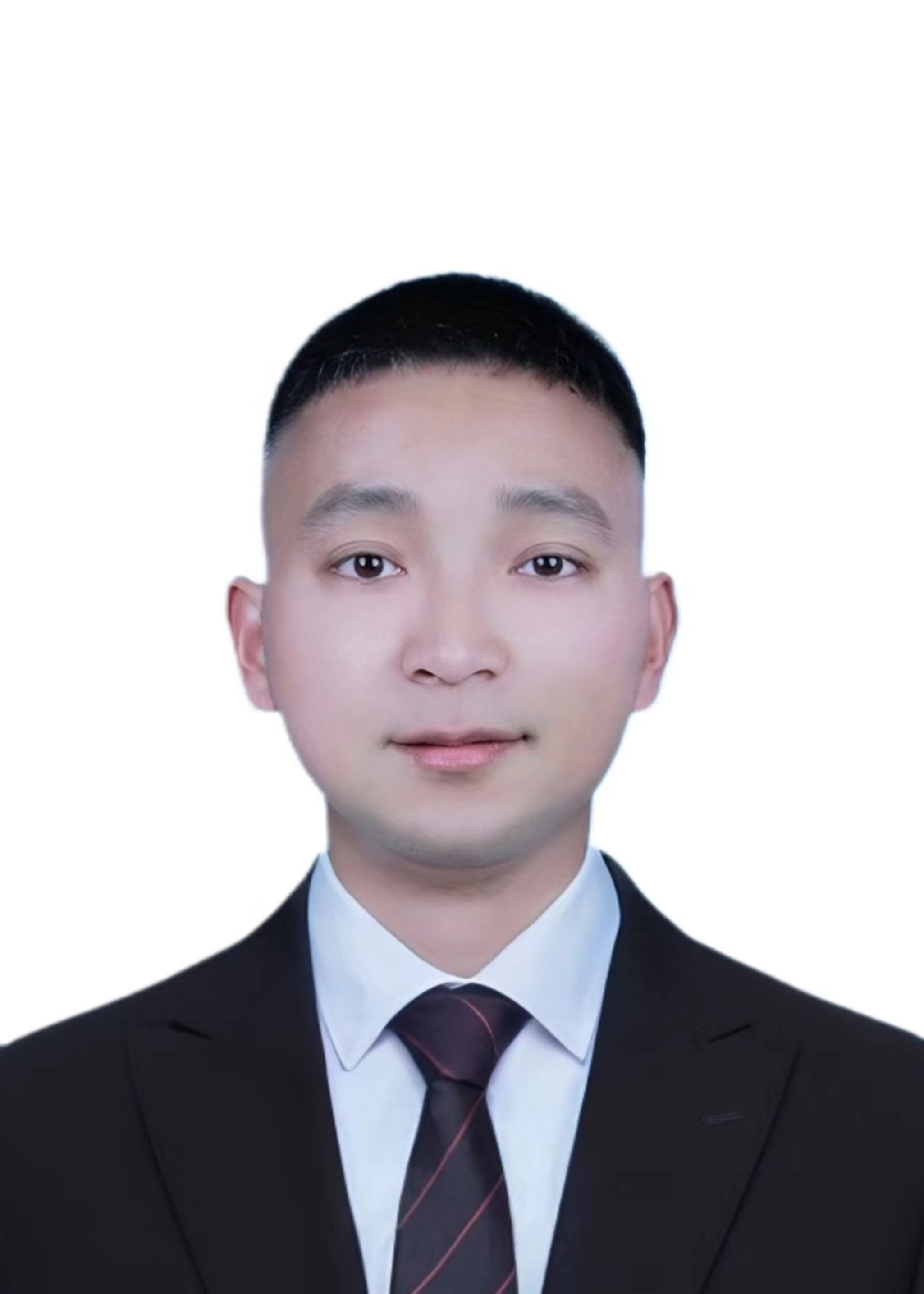}}]{Huaming Du}
received the Ph.D. degree in Southwestern University of Finance and Economics, Chengdu, China. He is a postdoctoral researcher with the Southwestern University of Finance and Economics, China.  His research interests include Large Language Model, Causal Inference, and graph representation learning. He has published papers in the prestigious refereed conferences and journals, such as KDD, IEEE TKDE, etc. \end{IEEEbiography}
\begin{IEEEbiography}[{\includegraphics[width=1in,height=1.25in,clip,keepaspectratio]{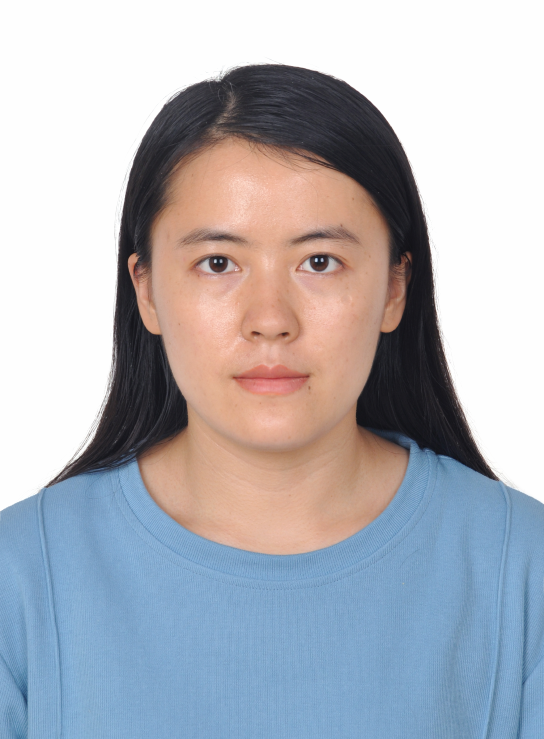}}]{Lei Yuan} is currently a PhD candidate in the School of Business Administration, Southwestern University of Finance and Economics. Her major research interests include causally-inspired machine learning and applications.
\end{IEEEbiography}
\begin{IEEEbiography}[{\includegraphics[width=1in,height=1.25in,clip,keepaspectratio]{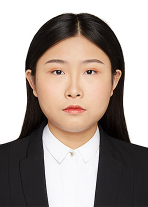}}]{Qing Yang} Qing Yang received the B.S. and M.S. degrees from Southwestern University of Finance and Economics in 2021 and 2024, respectively. Her research interests include enterprise risk forecasting.
\end{IEEEbiography}
\begin{IEEEbiography}[{\includegraphics[width=1in,height=1.25in,clip,keepaspectratio]{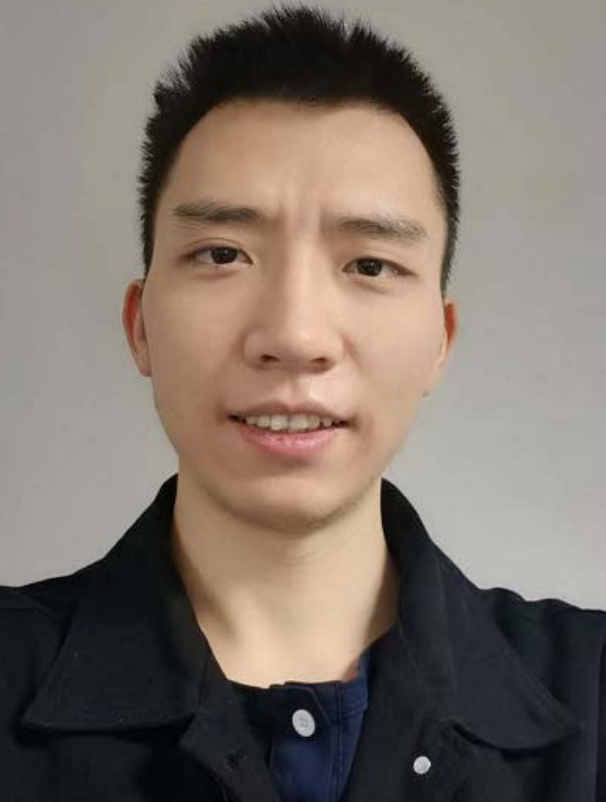}}]{Xingyan Chen}
received the Ph. D degree in computer technology from Beijing University of Posts and Telecommunications (BUPT), in 2021. He is currently a lecturer with the School of Economic Information Engineering, Southwestern University of Finance and Economics. He has published papers in well-archived international journals and proceedings, such as the IEEE TMC, IEEE TCSVT, IEEE TII, and IEEE INFOCOM, etc. His research interests include Multimedia Communications, Multi-agent RL.
\end{IEEEbiography}
\begin{IEEEbiography}[{\includegraphics[width=1in,height=1.25in,clip,keepaspectratio]{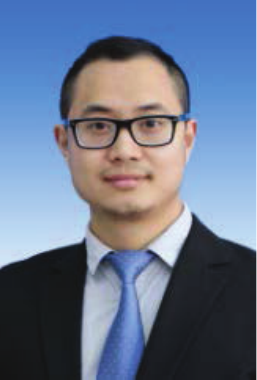}}]{Yu Zhao}
received the B.S. degree from Southwest Jiaotong University in 2006, and the M.S. and Ph.D. degrees from the Beijing University of Posts and Telecommunications in 2011 and 2017, respectively. He is currently an Associate Professor at Southwestern University of Finance and Economics. His current research interests include machine learning, natural language processing, knowledge graph, Fintech. He has authored more than 30 papers in top journals and conferences including IEEE TKDE, IEEE TNNLS, IEEE TMC, ACL, ICME, etc.
\end{IEEEbiography}
\begin{IEEEbiography}[{\includegraphics[width=1in,height=1.25in,clip,keepaspectratio]{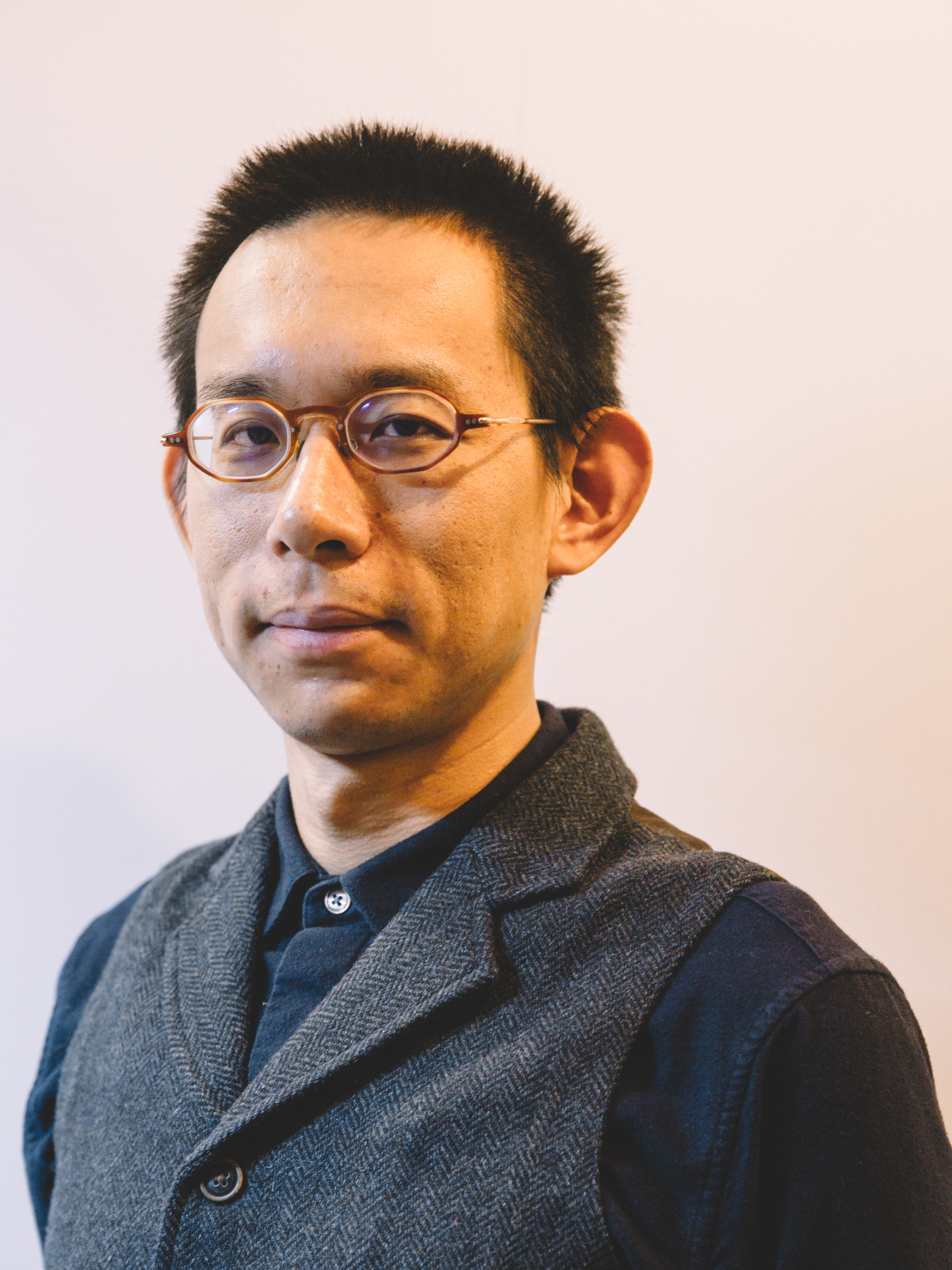}}]{Han Ji} is a seasoned fintech expert, previously served as CTO of Decent Galaxy Quantitative Fund,  former appointed expert by Wealth Management Association Qingdao (WMAQ), currently leading AntGroup’s Investment Research Technology Department.  He specializes in the integration and innovation of quantitative finance and AI technologies, with profound expertise in quantitative analysis, asset management, intelligent investment research, and LLM based multi-agent systems, having developed industry-leading solutions including the quantitative platform AntAlpha, AI-powered financial expert assistant Zhixiaozhu, and a open-source industrial-grade multi-agent framework agentUniverse. 
\end{IEEEbiography}
\begin{IEEEbiography}[{\includegraphics[width=1in,height=1.25in,clip,keepaspectratio]{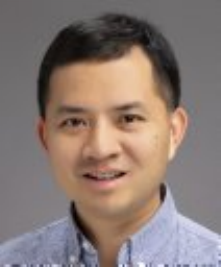}}]{Fuzhen Zhuang} received the PhD degree in computer science from the Institute of Computing Technology, Chinese Academy of Sciences. He is currently a full Professor in Institute of Artificial Intelligence, Beihang University. His research interests include Machine Learning and Data Mining, including Transfer Learning, Multi-task Learning, Multi-view Learning and Recommendation Systems. He has published more than 100 papers in the prestigious refereed conferences and journals, such as KDD, WWW, SIGIR, ICDE, IJCAI, AAAI, EMNLP, Nature Communications, IEEE TKDE, ACM TKDD, IEEE T-CYB, IEEE TNNLS, ACM TIST, etc.
\end{IEEEbiography}
\begin{IEEEbiography}[{\includegraphics[width=1in,height=1.25in,clip,keepaspectratio]{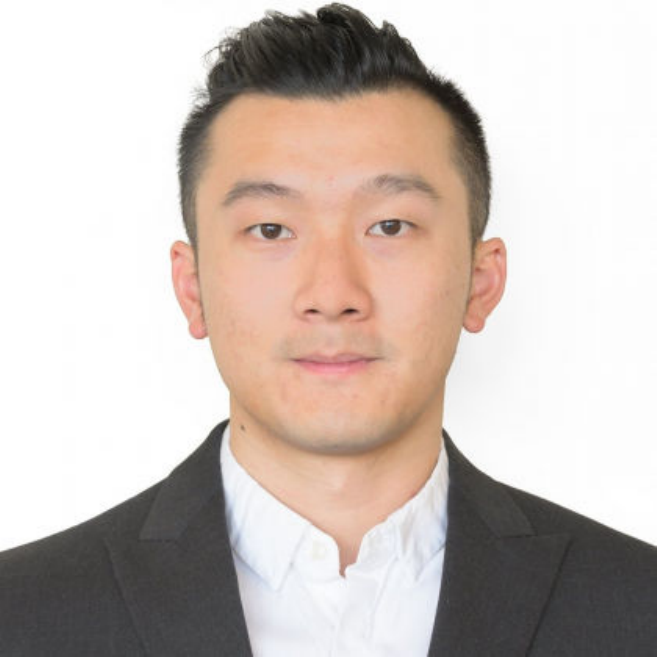}}]{Carl Yang} is an Assistant Professor of Computer Science at Emory University, jointly appointed at the Department of Biostatistics and Bioinformatics in the Rollins School of Public Health and the Center for Data Science in the Nell Hodgson Woodruff School of Nursing. He received his Ph.D. in Computer Science at University of Illinois, Urbana-Champaign in 2020, and B.Eng. in Computer Science and Engineering at Zhejiang University in 2014. His research interests span graph data mining, applied machine learning, knowledge graphs and federated learning, with applications in recommender systems, social networks, neuroscience and healthcare. Carl's research results have been published in 170+ peer-reviewed papers in top venues across data mining and biomedical informatics.
\end{IEEEbiography}
\begin{IEEEbiography}[{\includegraphics[width=1in,height=1.25in,clip,keepaspectratio]{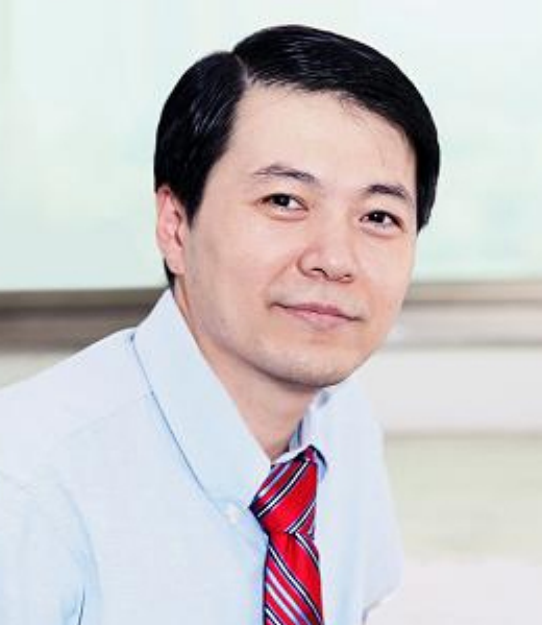}}]{Gang Kou} is a Distinguished Professor of Chang Jiang Scholars Program in Southwestern University of Finance and Economics, managing editor of International Journal of Information Technology \& Decision Making (SCI) and managing editor-in-chief of Financial Innovation (SSCI). He is also editors for other journals, such as: Decision Support Systems, and European Journal of Operational Research. Previously, he was a professor of School of Management and Economics, University of Electronic Science and Technology of China, and a research scientist in Thomson Co., R \& D. He received his Ph.D. in Information Technology from the College of Information Science \& Technology, Univ. of Nebraska at Omaha; Master degree in Dept of Computer Science, Univ. of Nebraska at Omaha; and B.S. degree in Department of Physics, Tsinghua University, China. He has published more than 100 papers in various peer-reviewed journals. Gang Kou’s h-index is 76 and his papers have been cited for more than 22000 times. He is listed as the Highly Cited Researcher by Clarivate Analytics (Web of Science).
\end{IEEEbiography}

\end{document}